\documentclass[epj]{svjour}
\usepackage{graphicx}

\def\mb#1{\setbox0=\hbox{$#1$}\kern-.025em\copy0\kern-\wd0
\kern-0.05em\copy0\kern-\wd0\kern-.025em\raise.0233em\box0}

\begin{document}
   \title{Relaxation of a test particle in systems with long-range interactions: diffusion coefficient and dynamical friction}

 \author{P.H. Chavanis}

\institute{ Laboratoire de Physique Th\'eorique, Universit\'e Paul
Sabatier, 118 route de Narbonne 31062 Toulouse, France\\
\email{chavanis@irsamc.ups-tlse.fr} }

\titlerunning{Test particle in systems with long-range interactions}

   \date{To be included later }

   \abstract{We study the relaxation of a test particle immersed
   in a bath of field particles interacting via weak long-range
   forces. To order $1/N$ in the $N\rightarrow +\infty$ limit, the
   velocity distribution of the test particle satisfies a
   Fokker-Planck equation whose form is related to the Landau and
   Lenard-Balescu equations in plasma physics. We provide explicit
   expressions for the diffusion coefficient and friction force in the
   case where the velocity distribution of the field particles is
   isotropic. We consider (i) various dimensions of space $d=3,2$ and
   $1$ (ii) a discret spectrum of masses among the particles (iii)
   different distributions of the bath including the Maxwell
   distribution of statistical equilibrium (thermal bath) and the step
   function (water bag). Specific applications are given for
   self-gravitating systems in three dimensions, Coulombian systems in
   two dimensions and for the HMF model in one dimension.
\PACS{
   {05.20.-y}{Classical statistical mechanics} \and
   {05.45.-a}{Nonlinear dynamics }} }

   \maketitle
%

\section{Introduction}
\label{sec_introduction}

Kinetic theories of many-particles systems are important to understand
the dynamical evolution of the system and to determine transport
properties. The first kinetic equation for a Hamiltonian $N$-body
system was derived by Boltzmann in his theory of gases
\cite{boltzmann}. In that case, the particles do not interact except
during strong collisions. Kinetic theories were later extended to the
case of particles in interaction by Landau
\cite{landau} in the case of plasmas and by Chandrasekhar
\cite{chandra} in the case of stellar systems. In this Introduction, 
we present a short historical review of kinetic equations with
gravitational or coulombian interaction, stressing the literature
before the sixties when these topics were developed. Additional
references can be found in the books of Balescu \cite{balescubook} and Saslaw \cite{saslaw}, or in the
review of Kandrup \cite{kandruprev}. The rest of the paper considers extensions of these
kinetic theories to new situations.

Landau \cite{landau} derived his kinetic equation by starting from the
Boltzmann equation and considering a weak deflection limit. Indeed,
for a Coulombian potential of interaction slowly decreasing with the
distance as $r^{-1}$, weak collisions are the most frequent ones. Each
encounter induces a {\it small} change in the velocity of a particle
but the cumulated effect of these encounters leads to a macroscopic
process of diffusion in velocity space. This treatment, which assumes
that the particles follow linear trajectories with constant velocity
in a first approximation, yields a logarithmic divergence of the
diffusion coefficient for both small and large impact parameters but
the equation can still be used successfully if appropriate cut-offs
are introduced. A natural lower cut-off, which is called the Landau
length, corresponds to the impact parameter leading to a deflection at
$90^{o}$. On the other hand, in a neutral plasma, the potential is
screened on a distance corresponding to the Debye
length. Phenomenologically, the Debye length provides an upper
cut-off. Later on, Lenard
\cite{lenard} and Balescu
\cite{balescu} developed a more precise (but also more formal) kinetic
theory that could take into account collective effects. This gives
rise to the inclusion of the dielectric function $|\epsilon({\bf
k},{\bf k}\cdot{\bf v})|^{2}$ in the denominator of the kinetic
equation. Physically, this means that the particles are ``dressed'' by
a polarization cloud. The original Landau equation, which ignores
collective effects, is recovered from the Lenard-Balescu equation when
$|\epsilon({\bf k},{\bf k}\cdot{\bf v})|^{2}=1$. However, with this
additional term, it is found that the logarithmic divergence at large
scales is now removed and that the Debye length is indeed the natural
upper lengthscale to consider.

In stellar dynamics, Chandrasekhar \cite{chandra} developed a kinetic
theory in order to determine the timescale of collisional
relaxation. He computed in particular the coefficients of diffusion
and friction (second and first moments of the velocity increments) by
considering the mean effect of a succession of two-body
encounters. Since this approach can take into account large
deflections, there is no divergence at small impact parameters and
the gravitational analogue of the Landau length appears
naturally.  However, this approach leads to a logarithmic divergence
at large scales that is more difficult to remove than in plasma
physics because of the absence of Debye shielding for the
gravitational force. In a series of papers, Chandrasekhar \& von
Neumann \cite{channeu} developed a completely stochastic formalism of
gravitational fluctuations and showed that the fluctuations of the
gravitational force are given by the Holtzmark distribution (a
particular L\'evy law) in which the nearest neighbor plays a prevalent
role. From these results, they argued that the logarithmic divergence
has to be cut-off at the interparticle distance. However, since the
interparticle distance is smaller than the Debye length, the same
arguments should also apply in plasma physics, which is not the
case. Therefore, the conclusions of Chandrasekhar \& von Neumann
\cite{channeu} are usually taken with circumspection. It is usually argued 
(e.g., Cohen {\it et al.} \cite{cohen}, Saslaw \cite{saslaw}, de Vega
\& Sanchez \cite{vega}) that the logarithmic divergence should
be cut-off at the physical size $R$ of the cluster for finite systems
or at the Jeans scale for infinite systems, since the Jeans length is
the presumable analogue of the Debye length in the present context
(see Kandrup \cite{kandruprev}). Chandrasekhar
\cite{chandraB} also developed a Brownian theory of stellar dynamics
and showed that, on a qualitative point of view, the results of
kinetic theory could be understood very simply in that framework. In
particular, he showed that the coefficients of diffusion and friction
are related to each other by an Einstein relation \cite{nice}. Later
on, Rosenbluth {\it et al.}
\cite{rosen} proposed a simplified derivation of the coefficients of
diffusion and friction for plasmas and stellar systems and,
substituting these expressions in the general form of the
Fokker-Planck equation, they obtained a kinetic equation which is at
the basis of the dynamics of stellar systems. They also provided
simplified expressions of this equation in the case of axial symmetry.

It is interesting to note that the previous authors did not point out
the link with the Landau equation and that modern textbooks of
astrophysics \cite{bt} usually derive the kinetic equation of stellar
dynamics from the Fokker-Planck equation by using the approach of
Chandrasekhar \cite{chandra} and Rosenbluth {\it et al.} 
\cite{rosen}. We will see that we
can equivalently obtain the kinetic equation of stellar dynamics from
the Landau equation. This alternative derivation can re-inforce the
connection between stellar systems and plasmas. We note, however, an
important difference between stellar dynamics and plasma
physics. Neutral plasmas are usually spatially homogeneous due to the
Debye shielding.  By contrast, stellar systems are
inhomogeneous. Therefore, the above-mentioned kinetic theories
developed in astrophysics rely on a {\it local approximation}. The
collision term is calculated as if the system were homogeneous or as
if the collisions could be treated as local. Then, the effect of
inhomogeneity is taken into account in the kinetic equation by
introducing an advective term (Vlasov term) in the left hand side
which descibes the evolution of the system due to mean-field
effects. The local approximation is supported by the stochastic
approach of Chandrasekhar \& Von Neumann \cite{channeu} showing the
preponderence of the nearest neighbor. However, this remains a
simplifying assumption which is not easily controllable. It is likely
that the logarithmic divergence at large scales comes from this
approximation. More recently, Kandrup \cite{kandrup1} derived a
generalized Landau equation by using projection operator
technics. This formal approach is interesting because it can take into
account effects of spatial inhomogeneity and memory which are
neglected in the previous approaches \footnote{Memory effects can be
important for gravitating systems because the temporal correlation
function of the force decreases algebraically as $t^{-1}$
\cite{chandra44}. However, this slow decay mainly results in
logarithmic divergences of the diffusion coefficient \cite{lee}, so
that the Landau equation can still be used successfully. Furthermore,
as shown in Saslaw \cite{saslaw}, non-Markovian effects are important
for long-range collisions but not for short-range collisions. Since
long-range collisions are more gentle than short-range collisions,
memory effects are somehow `washed-out' in the complete collision
process. Memory effects can be important for other systems with
long-range interactions, like the HMF model (see below), close to the
critical point $T_{c}$ because the timescale of the exponential decay
of the correlation function diverges as $T\rightarrow T_{c}$
\cite{bouchet,cvb,hb1,hb2}.}. It clearly shows which approximations are
needed in order to recover the Landau equation. However, the
generalized Landau equation remains extremely complicated for
practical purposes.

Until now, the kinetic theories of systems with long-range
interactions have been essentially developed for 3D systems with
Coulombian or Newtonian potential. The main object of this paper is to
extend these theories to other dimensions of space $d=2$ and $d=1$ and
for a wide class of potentials of interaction. This generalization has
been initiated in Chavanis \cite{hb1,hb2}. It is shown that for
systems with weak long-range interactions, the Landau and
Lenard-Balescu equations describe the collisional dynamics of the
system to order $1/N$ in a proper thermodynamic limit $N\rightarrow
+\infty$. Therefore, the collisional relaxation time scales in general
as $t_{R}\sim Nt_{D}$, where $t_{D}$ is the dynamical time. For
$N\rightarrow +\infty$ or $t\ll t_{R}$, the collision term is
negligible and we obtain the (mean-field) Vlasov equation.
In Ref. \cite{hb2}, we have introduced a
Fokker-Planck equation which describes the evolution of a test
particle in a bath of field particles. We have obtained analytical
expressions of the diffusion coefficent and friction force in the case
of a Maxwellian distribution of field particles (thermal bath). We shall obtain here generalizations of these results. 

In the
case of a Newtonian or Coulombian potential of interaction in $d=2$
the diffusion coefficient diverges {\it linearly} at large scales so
that an upper cut-off (related to the Debye length in the case of a
plasma) must be introduced. We have suggested in
\cite{hb2} that this divergence would be removed if we use the
Lenard-Balescu kinetic equation taking into account collective effects
as in the 3D case. This point will be further explored in the present
paper. The kinetic theory of Coulombian interactions in $d=2$ has been
studied independently by Benedetti {\it et al.} \cite{ben} using a
different approach. They calculated the friction force experienced by
a test particle by considering the effect of a succession of binary
encounters.  The calculations are difficult because they involve the
differential cross section which has not an explicit form in
$d=2$. They however managed to obtain asymptotic results for small and
large velocities in the case where the distribution of the field
particles is a step function. Our approach, based on the Landau or
Lenard-Balescu equation in $d=2$, does not require the expression of
the differential cross section, just the Fourier transform of the
potential of interaction. In this paper, we will calculate the
diffusion coefficient and the friction force in the case where the
distribution of the field particles is a step function and obtain
results that are compatible with those of
\cite{ben}. We will also discuss the stochastic
properties of the fluctuating force created by a 2D Coulombian system
by making a parallel with the approach of Chandrasekhar \& von Neumann
\cite{channeu} for the gravitational force in $d=3$. 

In $d=1$, the Landau and Lenard-Balescu collision terms cancel out
indicating that the collisional evolution of the system as a whole is
due to higher order correlations in the $1/N$ expansion
\cite{hb2}. However, if we use this kinetic theory to describe the
evolution of a test particle in a bath of field particles with any
(stable) steady distribution of the Vlasov equation, we obtain a
Fokker-Planck equation in which the diffusion coefficient and friction
force can be easily calculated. This type of Fokker-Planck equations
has been studied in \cite{bouchet,bd,cvb,cl} in connection with the HMF
model. A kinetic theory of the HMF model has been developed in Bouchet
\cite{bouchet} and Bouchet \& Dauxois \cite{bd} by analyzing the
stochastic process of fluctuations and calculating the first and
second moments of the velocity increment $\langle \Delta v\rangle$ and
$\langle (\Delta v)^{2}\rangle$. This is a particular case, for a
cosine potential of interaction in one dimension, of the general
Fokker-Planck approach developed in plasma physics (see Chap. 8 of
Ichimaru \cite{ichimaru}). These authors used this kinetic theory to
determine the velocity correlation function and found an algebraic
decay (see also \cite{marksteiner,lutz}) which is consistent with
direct numerical simulations of the $N$-body system. A kinetic theory
of the HMF model was developed independently by Chavanis (see
discussion in \cite{cvb}) by using the projection operator formalism.
This approach does not take into account collective effects but these
terms can be obtained from the Lenard-Balescu theory
\cite{hb2}. Chavanis \& Lemou \cite{cl} used this kinetic theory to
study the relaxation of the distribution function tails and showed
that it has a front structure.  Non-ideal effects in the kinetic
theory (non-Markovianity, spatial inhomogeneity,...) have been further
discussed in
\cite{curious,angle}. In the present paper, we shall develop a kinetic
theory of homogeneous one-dimensional systems with an arbitrary form
of weak long-range potential of interaction, starting directly from
the general Fokker-Planck equation given in \cite{hb2}, which takes
into account collective effects. We shall determine the diffusion
coefficient and analyze the effects of a mass distribution among the
particles.

We note finally that kinetic theories of systems with long-range
interactions have also been developed for point vortices in
two-dimensional hydrodynamics and non-neutral plasmas confined by a
magnetic field. Their form is related to, but different from, the
Landau and Lenard-Balescu equations. The intrinsic reason is that
point vortices do not have inertia contrary to electric charges and
stars. Hence, the coordinates $x$ and $y$ are canonically
conjugate. Kinetic equations have been derived independently by Dubin
\& O'Neil
\cite{dn,dubin} from the Klimontovich approach and by Chavanis
\cite{kin} from projection operator technics. 
On the other hand, by using an analogy with stellar dynamics and
Brownian theory, a Fokker-Planck equation describing the stochastic
evolution of a test vortex in a bath of field vortices at equilibrium
was derived in \cite{preR,kin}.  The diffusion coefficient and the drift
term are related to each other by an appropriate Einstein relation and
they are inversely proportional to the local shear created by the
field vortices. The statistics of the velocity fluctuations arising
from a random distribution of point vortices has been investigated in
Chavanis \& Sire \cite{cs,csfluid} by using an approach similar to
that developed by Chandrasekhar \& von Neumann
\cite{channeu} for the gravitational force. The analogy between
stellar systems and 2D vortices is discussed in \cite{houches}.

In this paper, we shall complete previous investigations on the
kinetic theory of systems with weak long-range interactions by
developing a general formalism valid for a large class of potentials of
interaction in $d$ dimensions and for multi-components systems. In
Sec. \ref{sec_landau}, we recall the Lenard-Balescu and Landau
equations describing the evolution of the distribution function (DF)
of a system of particles in interaction. In Sec. \ref{sec_fp}, we
introduce the Fokker-Planck equation describing the evolution of a
test particle immersed in a bath of field particles. We give the
general expressions of the coefficients of diffusion and friction. In
Sec. \ref{sec_isotropic}, we restrict ourselves to isotropic
distribution functions and derive the simplified form of the
Fokker-Planck equation. We show how its stationary solutions are
related to the distribution of the bath and we note that the test
particle does not in general relax towards the distribution of the
bath except (i) if it is Maxwellian (thermal bath) (ii) in $d=1$ for
single-species particles. In Sec. \ref{sec_einstein}, we consider the
case where the field particles have a Maxwellian distribution. We find
that the Fokker-Planck equation becomes similar to the Kramers
equation but the diffusion is anisotropic and depends on the velocity
of the test particle. The coefficients of diffusion and friction are
related by an Einstein relation. We derive its general form for a
multi-components system. In Sec. \ref{sec_diffusion}, we derive the
explicit expression of the diffusion coefficient of a test particle in
a thermal bath in $d=1,2,3$. In Sec. \ref{sec_time}, we estimate the
relaxation time of the test particle to the Maxwellian distribution
(thermalization) and show that it scales as $N t_{D}$. We also
investigate the effect of mass seggregation on the relaxation
time. Specific applications are given for gravitational systems and for the
HMF model in Sec. \ref{sec_examples}.  In Sec. \ref{sec_d3}, we
consider the case $d=3$. We show that the results obtained by
Rosenbluth {\it et al.}
\cite{rosen} can be alternatively obtained from the Landau
equation. This has not been noted by the previous authors and this can
re-inforce the connection between plasma physics and stellar
dynamics. As an application of the Rosenbluth potentials, we compute
the diffusion coefficient and the friction force in the case where the
distribution of the bath is a step function. In Sec. \ref{sec_d2}, we
consider the case $d=2$. We obtain the general expressions of the
diffusion coefficient and friction force valid for any isotropic
distribution of the field particles in terms of integrals of Bessel
functions.  As an illustration, we compute them for the Coulombian
interaction when the distribution of the bath is a step function and
we obtain results similar to those of \cite{ben} up to a factor
$2(\pi-2)/\pi$ that may be related to the unknown large scale
cut-off. In Sec. \ref{sec_statistics}, we analyze the statistics of
fluctuations of the Coulombian force in $d=2$. We show that it is
given by a marginal Gaussian distribution intermediate between normal
and L\'evy laws. In particular, it presents an algebraic tail scaling
as $\sim F^{-4}$. This is analogous to the distribution of the
velocity created by a gas of point vortices in two-dimensions
\cite{cs,csfluid}. We also discuss the origin of the linear divergence of the
diffusion coefficient for Coulombian systems in $d=2$ and how this
can be cured by considering collective effects. Finally, in
Sec. \ref{sec_d1} we provide the general form of the Fokker-Planck
equation in $d=1$ and give explicit expressions of the diffusion
coefficient and friction force for any (stable) steady distribution of
the bath and in the case where the test particle has not necessarily
the same mass as the field particles.

\section{General results}
\label{sec_general}

\subsection{Evolution of the system as a whole: the Landau and Lenard-Balescu equations}
\label{sec_landau}

We consider a system of particles with long-range interactions whose
dynamics is described by the Hamiltonian equations
\begin{eqnarray}
\label{landau1}
m_{i}\frac{d{\bf r}_{i}}{dt}=\frac{\partial H}{\partial {\bf v}_{i}}, \qquad m_{i}\frac{d{\bf v}_{i}}{dt}=-\frac{\partial H}{\partial {\bf r}_{i}},
\end{eqnarray}
where
\begin{eqnarray}
\label{landau2}
H=\sum_{i}^{N}m_{i}\frac{v_{i}^{2}}{2}+\sum_{i<j}\gamma_{i}\gamma_{j} u({\bf r}_{i}-{\bf r}_{j}),
\end{eqnarray}
where $u_{ij}=u({\bf r}_{i}-{\bf r}_{j})$ is a binary potential of
interaction depending only on the absolute distance between
particles. We assume that there exists $X$ species of particles
$\lbrace m_{i},\gamma_{i}\rbrace_{i=1,X}$ and we denote $f_{i}({\bf
r},{\bf v},t)$ the distribution function of particles of species $i$
normalized such that $\int f_{i}d{\bf r}d{\bf v}=N_{i}m_{i}$ gives the
total mass of particles of species $i$. We consider homogeneous
systems and we assume that the potential of interaction is long-range
and of weak amplitude. Then, the evolution of the distribution
function of species $i$ is given by the Lenard-Balescu equation
\begin{eqnarray}
\label{landau3}
\frac{\partial f_{i}}{\partial t}=\pi (2\pi)^{d}\frac{\partial}{\partial v^{\mu}}\int d{\bf v}'d{\bf k} k^{\mu}k^{\nu}\frac{\hat{u}(k)^{2}}{|\epsilon({\bf k},{\bf k}\cdot{\bf v})|^{2}}\delta({\bf k}\cdot {\bf u})\nonumber\\
\times \left (\frac{\gamma_{i}}{m_{i}}\right )^{2}\sum_{j} \left (\frac{\gamma_{j}}{m_{j}}\right )^{2}\left (m_{j}f'_{j}\frac{\partial f_{i}}{\partial v^{\nu}}-m_{i}f_{i} \frac{\partial f'_{j}}{\partial v^{'\nu}}\right )
\end{eqnarray} 
with the dielectric function
\begin{eqnarray}
\label{landau4}
\epsilon({\bf k},\omega)=1+(2\pi)^{d}\hat{u}(k)\sum_{j}\left (\frac{\gamma_{j}}{m_{j}}\right )^{2}\int \frac{{\bf k}\cdot \frac{\partial f_{j}}{\partial {\bf v}}}{\omega-{\bf k}\cdot {\bf v}}d{\bf v}.\nonumber\\
\end{eqnarray} 
We have introduced the abbreviations $f_{i}=f_{i}({\bf v},t)$ and
$f'_{i}=f_{i}({\bf v}',t)$. Furthermore, ${\bf u}={\bf v}-{\bf v}'$ is
the relative velocity between two particles. It is also implicitly
understood that there is summation over repeated greek indices which
denote the cartesian coordinates of the vectors.

For systems with weak long-range interactions, the Lenard-Balescu
equation gives the correction of order $1/N$ to the Vlasov equation
which is recovered for $N\rightarrow +\infty$ (for homogeneous systems
the Vlasov equation simply reduces to $\partial f/\partial t=0$). The
proper thermodynamic limit is defined in \cite{hb1,hb2}. It is such
that the amplitude of the potential of interaction scales as $u\sim
1/N$ (weak coupling) while the energy per particle $E/N$, the inverse
temperature $\beta$ and the volume $V$ are of the order unity.  Noting
that $u^{2}\sim 1/N^{2}$ and $f\sim N$, we find that the
Lenard-Balescu collision term is of order $1/N$, i.e. $\partial
f/\partial t=\frac{1}{N}Q(f)$ with $Q(f)\sim f$. In $d=3$ and $d=2$,
it is easy to show that the Lenard-Balescu equation relaxes toward the
Maxwellian
\begin{eqnarray}
\label{landau5}
f_{i}^{eq}=A_{i}e^{-\beta m_{i}\frac{v^{2}}{2}}.
\end{eqnarray}  
Therefore, finite $N$ effects select the Maxwell distribution among
all possible stationary solutions of the Vlasov equation. The
collisional relaxation time scales as $t_{R}\sim N t_{D}$ where
$t_{D}$ is a dynamical time (which can be taken of order unity). For
Newtonian interactions in $d=3$, there is a logarithmic correction
$\ln\Lambda\sim \ln N$ so that the collisional relaxation time scales
as $t_{R}\sim (N/\ln N) t_{D}$. Alternatively, for $d=1$, the
Lenard-Balescu operator cancels out so that the collisional evolution
is due to terms of higher order in $1/N$. This implies that the
collisional relaxation time of the system as a whole scales as
$t_{R}\sim N^{\delta} t_{D}$ with $\delta>1$.

If we neglect collective effects and take $|\epsilon({\bf k},{\bf k}\cdot{\bf v})|^{2}=1$ we obtain the Landau equation
\begin{eqnarray}
\label{landau6}
\frac{\partial f_{i}}{\partial t}=\pi (2\pi)^{d}\frac{\partial}{\partial v^{\mu}}\int d{\bf v}'d{\bf k} k^{\mu}k^{\nu}{\hat{u}(k)^{2}}\delta({\bf k}\cdot {\bf u})\nonumber\\
\times \left (\frac{\gamma_{i}}{m_{i}}\right )^{2}\sum_{j} \left (\frac{\gamma_{j}}{m_{j}}\right )^{2}\left (m_{j}f'_{j}\frac{\partial f_{i}}{\partial v^{\nu}}-m_{i}f_{i} \frac{\partial f'_{j}}{\partial v^{'\nu}}\right )
\end{eqnarray} 
as an approximation of the Lenard-Balescu equation. An important
remark for the following is that the expression (\ref{landau6}) of the
Landau equation does not involve the differential cross section of the
interaction, but only the Fourier transform of the potential
\footnote{The Fourier transform of the potential is just the Born
approximation to the scattering amplitude. Therefore, the assumptions
that are made to obtain the Landau equation amount to a first order
perturbative approximation for the elastic two-body differential cross
section (I thank one of the referees for this remark).}. Furthermore,
as discussed after Eq. (\ref{rosen2}), the potential of interaction
only fixes the timescale of relaxation so that the {\it structure} of
the Landau equation does not depend on the potential.  This is a
consequence of the fact that the amplitude of the interaction is very
small so that a weak deflection approximation is appropriate and
yields results that are relatively independent on the precise form of
the potential of interaction.

The Landau equation (\ref{landau6}) can be derived from the Boltzmann
\cite{landau} or from the Fokker-Planck equation \cite{chandra,rosen}
by using a model of binary collisions. The Lenard-Balescu equation
(\ref{landau3}) can be obtained from the Liouville equation by using
iterative procedures and diagrammatic methods \cite{balescu}. It can
also be obtained from the BBGKY hierarchy \cite{ichimaru} or from the
Klimontovich equation \cite{pitaevskii} by using approximations which
amount to neglecting some correlations. This can be justified
perturbatively when we consider an expansion of the equations of the
problem in terms of the small parameter $\sim 1/N$ (weak coupling
limit). These derivations remain valid for other types of potentials
with weak amplitude and other dimensions of space so that
Eqs. (\ref{landau3}) and (\ref{landau6}) have a larger domain of
validity than plasma physics and stellar dynamics. These equations
will be the starting point of our analysis.

\subsection{Evolution of a test particle in a bath: the Fokker-Planck equation}
\label{sec_fp}

We now consider the evolution of a test particle $(m,\gamma)$ in a
bath of field particles with prescribed distribution $f_{j}({\bf v})$
for each species $(m_{j},\gamma_{j})$. It has been established in
plasma physics (see, e.g.  \cite{ichimaru}) that the evolution of the
velocity distribution $P({\bf v},t)$ of the test particle is governed
by a Fokker-Planck equation that takes a form similar to the
Lenard-Balescu equation (\ref{landau1}) provided that we replace
$f_{j}'({\bf v},t)$ by the {\it prescribed} distribution $f_{j}({\bf
v})$ of the bath. This general Fokker-Planck equation
\begin{eqnarray}
\label{fp1}
\frac{\partial P}{\partial t}=\pi (2\pi)^{d}\frac{\partial}{\partial v^{\mu}}\int d{\bf v}'d{\bf k} k^{\mu}k^{\nu}\frac{\hat{u}(k)^{2}}{|\epsilon({\bf k},{\bf k}\cdot{\bf v})|^{2}}\delta({\bf k}\cdot {\bf u})\nonumber\\
\times \left (\frac{\gamma}{m}\right )^{2}\sum_{j} \left (\frac{\gamma_{j}}{m_{j}}\right )^{2}\left (m_{j}f'_{j}\frac{\partial P}{\partial v^{\nu}}-m P \frac{\partial f'_{j}}{\partial v^{'\nu}}\right ),
\end{eqnarray} 
will be our main object of interest.  In the following, in order to
simplify the expressions, we shall take $\gamma_{i}=m_{i}$ (like for
the gravitational interaction) but the generalization of our results
for $\gamma_{i}\neq m_{i}$ is straightforward. Furthermore, for $d=3$
and $d=2$ we shall make the Landau approximation $|\epsilon({\bf
k},{\bf k}\cdot{\bf v})|^{2}=1$ (some results that relax the Landau approximation are given in \cite{hb2} for a thermal bath). In that case, the
Fokker-Planck equation for the test particle becomes
\begin{eqnarray}
\label{fp2}
\frac{\partial P}{\partial t}=\frac{\partial}{\partial v^{\mu}}\sum_{j}\int K^{\mu\nu}\left (m_{j}f'_{j}\frac{\partial P}{\partial v^{\nu}}-m P\frac{\partial f'_{j}}{\partial v^{'\nu}}\right )d{\bf v}'
\end{eqnarray}
where 
\begin{eqnarray}
\label{fp3}
K^{\mu\nu}=\pi (2\pi)^{d}\int k^{\mu}k^{\nu}\hat{u}(k)^{2}\delta({\bf k}\cdot {\bf u})d{\bf k}.
\end{eqnarray}
We introduce the diffusion tensor and the friction term:
\begin{eqnarray}
\label{fp4}
D^{\mu\nu}=\sum_{j}\int K^{\mu\nu} m_{j}f'_{j}d{\bf v}'\equiv \sum_{j}D^{\mu\nu}(j\rightarrow 0),
\end{eqnarray}
\begin{eqnarray}
\label{fp5}
\eta^{\mu}=-m\sum_{j}\int K^{\mu\nu} \frac{\partial f'_{j}}{\partial v'^{\nu}}d{\bf v}'\equiv \sum_{j}\eta^{\mu}(j\rightarrow 0),
\end{eqnarray}
where $D^{\mu\nu}(j\rightarrow 0)$ and $\eta^{\mu}(j\rightarrow 0)$
are the diffusion and the friction caused by species $j$ on the test
particle denoted $0$. With these notations,  Eq. (\ref{fp2}) can be rewritten
\begin{eqnarray}
\label{fp6}
\frac{\partial P}{\partial t}=\frac{\partial}{\partial v^{\mu}}\left\lbrack D^{\mu\nu}\frac{\partial P}{\partial v^{\nu}}+P\eta^{\mu}\right\rbrack.
\end{eqnarray}
Comparing with the general expression of the Fokker-Planck equation
\begin{eqnarray}
\label{fp7}
\frac{\partial P}{\partial t}=\frac{1}{2}\frac{\partial^{2}}{\partial v^{\mu}\partial v^{\nu}}\left (\frac{\langle \Delta v^{\mu}\Delta v^{\nu}\rangle}{\Delta t}P\right )+\frac{\partial}{\partial v^{\mu}}\left (\frac{\langle \Delta v^{\mu}\rangle}{\Delta t}P\right ),\nonumber\\
\end{eqnarray}
we find that
\begin{eqnarray}
\label{fp8}
\frac{\langle \Delta v^{\mu}\Delta v^{\nu}\rangle}{\Delta t}=2D^{\mu\nu}
\end{eqnarray}
\begin{eqnarray}
\label{fp9}
\frac{\langle \Delta v^{\mu}\rangle}{\Delta t}=\eta^{\mu}-\frac{\partial D^{\mu\nu}}{\partial v^{\nu}}.
\end{eqnarray}
Now, using the fact that $K^{\mu\nu}$ depends only on the relative velocity ${\bf u}={\bf v}-{\bf v}'$ and using an integration by parts, we get
\begin{eqnarray}
\label{fp10}
\frac{\partial D^{\mu\nu}}{\partial v^{\nu}}=\sum_{j}\int \frac{\partial K^{\mu\nu}}{\partial v^{\nu}} m_{j}f'_{j}d{\bf v}'\nonumber\\
=-\sum_{j}\int \frac{\partial K^{\mu\nu}}{\partial v^{'\nu}}
m_{j}f'_{j}d{\bf v}' =\sum_{j}\int K^{\mu\nu} m_{j}\frac{\partial
f'_{j}}{\partial v^{'\nu}}d{\bf v}'.\nonumber\\
\end{eqnarray}
Therefore, the first moment of the velocity increment can be rewritten
\begin{eqnarray}
\label{fp11}
\frac{\langle \Delta v^{\mu}\rangle}{\Delta t}=-\sum_{j}\int K^{\mu\nu}(m+m_{j})\frac{\partial
f'_{j}}{\partial v^{'\nu}}d{\bf
v}'\nonumber\\
=-\sum_{j}\frac{m+m_{j}}{m}\eta^{\mu}(j\rightarrow 0).
\end{eqnarray}
We note that because of the velocity dependence of the diffusion
tensor, the friction terms $\langle \Delta {\bf v}\rangle/{\Delta t}$
and ${\mb \eta}$ do not coincide. In particular, for equal mass
particles, there is a factor $2$ between them as noted previously
\cite{genlandau}. On the other hand, if the field particles have
a mass $m_{f}$ different from the mass $m$ of the test particle, the
multiplicative factor is $(m+m_{f})/m$. In this respect, we note that
the frictional force $\langle {\bf F}_{\rm fr}\rangle$ calculated by
Kandrup \cite{kandrup} with his linear response theory is ${\mb
\eta}$, not $\langle \Delta {\bf v}\rangle/{\Delta t}$. This explains
why it differs from the calculation of Chandrasekhar \cite{chandra} by
a factor $(m+m_{f})/m$ (see \cite{kandrup}, pp. 446). The frictional
force ${\mb \eta}$ is an important quantity by itself because it is
the quantity which naturally appears in the {\it symmetrical} form of
the Landau-Lenard-Balescu collision term where the diffusion
coefficient is placed between the two derivatives: $\partial_{\mu}
D^{\mu\nu}\partial_{\nu}$, see Eqs. (\ref{landau3}), (\ref{landau6}) 
and (\ref{fp6}).

\subsection{The isotropic Fokker-Planck equation}
\label{sec_isotropic}

If the distribution function of the field particles is isotropic, i.e. $f_{j}({\bf v})=f_{j}(v)$, we can write the diffusion tensor in the form
\begin{eqnarray}
\label{iso1}
D^{\mu\nu}=\left (D_{\|}-\frac{1}{d-1}D_{\perp}\right)\frac{v^{\mu}v^{\nu}}{v^{2}}+\frac{1}{d-1}D_{\perp}\delta^{\mu\nu},
\end{eqnarray}
where $D_{\|}(v)$ and $D_{\perp}(v)$ are the diffusion coefficients in
the directions parallel and perpendicular to the direction of the test
particle ${\bf v}$. On the other hand, starting from Eqs. (\ref{fp4})
and (\ref{fp5}) and using the same type of calculation as in
Eq. (\ref{fp10}), we find that the friction term can be  generally written
\begin{eqnarray}
\label{iso2}
\eta^{\mu}=-m\sum_{j}\frac{1}{m_{j}}\frac{\partial}{\partial v^{\nu}}D^{\mu\nu}(j\rightarrow 0).
\end{eqnarray}
For an isotropic distribution of the field particles, using
Eq. (\ref{iso1}), we find that the friction term is parallel to the
velocity of the test particle, i.e.  ${\mb\eta}=\eta {\bf v}/v$. The
amplitude of the friction vector is given by
\begin{eqnarray}
\label{iso3}
\eta=-m\sum_{j}\frac{1}{m_{j}}\left\lbrack \frac{dD_{\|}}{dv}+\frac{d-1}{v}\left (D_{\|}-\frac{D_{\perp}}{d-1}\right )\right\rbrack(j\rightarrow 0).\nonumber\\
\end{eqnarray}

If the velocity distribution of the test particle is itself isotropic,
i.e. $P({\bf v},t)=P(v,t)$, we can rewrite the Fokker-Planck equation
(\ref{fp6}) in the form
\begin{eqnarray}
\label{iso4}
\frac{\partial P}{\partial t}=\frac{1}{v^{d-1}}\frac{\partial}{\partial v}\left \lbrack v^{d-1}\left (D_{\|}\frac{\partial P}{\partial v}+P\eta\right )\right \rbrack
\end{eqnarray}
where we have used
\begin{eqnarray}
\label{iso5}
D^{\mu\nu}v^{\nu}=(D_{\|}-\frac{1}{d-1}D_{\perp})v^{\mu}+\frac{1}{d-1}D_{\perp}v^{\mu}=D_{\|}v^{\mu}.\nonumber\\
\end{eqnarray}
Equation (\ref{iso4})  can also be written as
\begin{eqnarray}
\label{iso6}
\frac{\partial P}{\partial t}=\frac{1}{v^{d-1}}\frac{\partial}{\partial v}\left \lbrack v^{d-1}D_{\|}(v)\left (\frac{\partial P}{\partial v}+P\frac{dU}{dv}\right )\right \rbrack
\end{eqnarray}
where we have introduced the effective potential
\begin{eqnarray}
\label{iso7}
U(v)=\int^{v}\frac{\eta(v')}{D_{\|}(v')}dv'.
\end{eqnarray}
Equation (\ref{iso6}) relaxes towards a stationary solution of the form
\begin{eqnarray}
\label{iso8}
P^{eq}(v)=Ae^{-U(v)},
\end{eqnarray}
provided that this distribution is normalizable.  The evolution of the
high velocity tail of the distribution function that is  solution of a
Fokker-Planck equation of the general form (\ref{iso6}) has been
studied in \cite{cl}.

Consider single species systems. For $d=2,3$, the {\it
only} stationary solution of the Lenard-Balescu equation
(\ref{landau3}) is the Maxwellian (\ref{landau5}). This implies that
the stationary solution of the Fokker-Planck equation (\ref{fp1}) will
be equal to the distribution of the bath $f({\bf v})$ {\it only} if
this distribution is the Maxwellian. Otherwise, $P^{eq}(v)$ is not
equal to the distribution $f({\bf v})$ of the bath. We shall give an
explicit example in Sec. \ref{sec_waterbag}. By contrast, for $d=1$,
the Lenard-Balescu operator cancels out for {\it any}
distribution. This implies that the stationary solution of the
Fokker-Planck equation (\ref{fp1}) is {\it always} equal to the
distribution of the bath $f({\bf v})$, even if it is not the
Maxwellian (see Sec. \ref{sec_d1}).

\subsection{The Einstein relation for isothermal systems}
\label{sec_einstein}

If the velocity distribution of the field particles is the Maxwellian (thermal bath)
\begin{eqnarray}
\label{einstein1}
f_{j}=A_{j}e^{-\beta m_{j}\frac{v^{2}}{2}},
\end{eqnarray}  
then 
\begin{eqnarray}
\frac{\partial f_{j}'}{\partial v^{'\nu}}=-\beta m_{j}f_{j}'v^{'\nu}.
\label{einstein2}
\end{eqnarray} 
Substituting this relation in Eq. (\ref{fp5}) and using the relation
\begin{eqnarray}
K^{\mu\nu}v^{'\nu}=K^{\mu\nu}(v^{\nu}-u^{\nu})=K^{\mu\nu}v^{\nu},
\label{einstein3}
\end{eqnarray}
which results from the identity $K^{\mu\nu}u^{\nu}=0$ ($K^{\mu\nu}$ is
the projector in the direction perpendicular to ${\bf u}$), we get
\begin{eqnarray}
\eta^{\mu}=\beta m D^{\mu\nu}v^{\nu}.
\label{einstein4}
\end{eqnarray} 
This  can be viewed as the general expression  of the Einstein relation in our
context (this is the most general relation that Eq. (\ref{fp6}) must satisfy in order to admit the Maxwell distribution as a stationary state). In that case, the Fokker-Planck equation (\ref{fp6}) can be written in a form similar to the Kramers equation \cite{risken}, but with an anisotropic tensor depending on the velocity of the test particle:
\begin{eqnarray}
\label{einstein5}
\frac{\partial P}{\partial t}=\frac{\partial}{\partial v^{\mu}}\left\lbrack D^{\mu\nu}(v)\left (\frac{\partial P}{\partial v^{\nu}}+\beta m P v^{\nu}\right )\right\rbrack.
\end{eqnarray}
The stationary solution of this equation is the Maxwellian
\begin{eqnarray}
\label{einstein6}
P^{eq}(v)=A e^{-\beta m \frac{v^{2}}{2}}.
\end{eqnarray}
Note also that according to Eq. (\ref{iso5}) we have
\begin{eqnarray}
{\mb \eta}=\beta m D_{\|}{\bf v}.
\label{einstein7}
\end{eqnarray} 
Therefore, if we assume that the velocity distribution of the test particle is isotropic we get
\begin{eqnarray}
\label{einstein8}
\frac{\partial P}{\partial t}=\frac{1}{v^{d-1}}\frac{\partial}{\partial v}\left \lbrack v^{d-1}D_{\|}(v)\left (\frac{\partial P}{\partial v}+\beta m P v\right )\right \rbrack.
\end{eqnarray}
If we momentarily neglect the velocity dependence of the diffusion coefficient, this Fokker-Planck equation can be obtained from a Langevin stochastic process of the form
\begin{eqnarray}
\label{einstein9}
\frac{d{\bf r}}{dt}={\bf v}, \quad \frac{d{\bf v}}{dt}=-\xi {\bf v}+\sqrt{2D_{\|}}{\bf R}(t),
\end{eqnarray}
where ${\bf R}(t)$ is a white noise satisfying $\langle {\bf
R}(t)\rangle=0$ and $\langle
{R}_{i}(t)R_{j}(t')\rangle=\delta_{ij}\delta(t-t')$ and the friction
coefficient is given by the Einstein relation $\xi=\beta m
D_{\|}$. This relation is {\it necessary} to obtain the Maxwellian
distribution at equilibrium and this is why it is rather independent
on the details of the microscopic model.

As indicated previously, since the diffusion coefficient depends on
the velocity, ${\mb \eta}$ is not exactly the friction force so that
Eq. (\ref{einstein7}) is not the proper form of Einstein
relation. Using Eq. (\ref{fp11}) we have
\begin{eqnarray}
\label{einstein10}
\frac{\langle \Delta {\bf v}\rangle}{\Delta t}=-\beta\sum_{j} (m+m_{j})D_{\|}(j\rightarrow 0){\bf v}.
\end{eqnarray}
If the field particles all have the same mass $m_{f}$ we obtain 
\begin{eqnarray}
\label{eisntein11}
\frac{\langle \Delta {\bf v}\rangle}{\Delta t}=-\beta (m+m_{f})D_{\|}{\bf v},
\end{eqnarray}
so that the proper friction coefficient is
\begin{eqnarray}
\label{einstein12}
\xi'=\beta (m+m_{f})D_{\|}.
\end{eqnarray}
This is the proper form of the Einstein relation in the present context. Note that it involves the sum of the mass of the test particle and of the field particles.  Using 
\begin{eqnarray}
\label{einstein13}
\frac{1}{2}m_{f}\langle v^{2}\rangle_{f}=\frac{d}{2}k_{B}T=\frac{d}{2\beta},
\end{eqnarray}
the Einstein relation (\ref{einstein12})  can be rewritten
\begin{eqnarray}
\label{einstein14}
\frac{\langle (\Delta v_{\|})^{2}\rangle}{\xi'\Delta t}=\frac{2}{d}\frac{m_{f}}{m+m_{f}}\langle v^{2}\rangle_{f}.
\end{eqnarray}

\subsection{The diffusion coefficient for isothermal systems}
\label{sec_diffusion}

Inserting the identity
\begin{eqnarray}
\delta({x})=\int_{-\infty}^{+\infty} e^{i{t}{x}}\frac{d{t}}{2\pi},
\label{diff1}
\end{eqnarray} 
in Eq. (\ref{fp3}) and performing the Fourier transform on ${\bf v}'$
of the Gaussian distribution (\ref{einstein1}) in Eq. (\ref{fp4}), we
find after another Fourier transform on $t$ that the diffusion tensor
can be expressed as
\begin{eqnarray}
D^{\mu\nu}=\pi (2\pi)^{d}\left (\frac{\beta}{2\pi}\right )^{1/2}\sum_{j}\rho_{j}m_{j}^{3/2}\nonumber\\
\times \int d{\bf k}\frac{k^{\mu}k^{\nu}}{k}\hat{u}(k)^{2}e^{-\beta m_{j}\frac{({\bf k}\cdot{\bf v})^{2}}{2k^{2}}}.
\label{diff2}
\end{eqnarray} 
This can be rewritten
\begin{eqnarray}
D^{\mu\nu}=\pi (2\pi)^{d}\left (\frac{\beta}{2\pi}\right )^{1/2}\sum_{j}\rho_{j}m_{j}^{3/2}\nonumber\\
\times\int_{0}^{+\infty} k^{d} \hat{u}(k)^{2} d{k} \ G^{\mu\nu}\left( \sqrt{\frac{\beta m_{j}}{2}}v\right ),
\label{diff3}
\end{eqnarray}
where 
\begin{eqnarray}
G^{\mu\nu}(x)=\int d{\bf \hat{k}}\ \hat{k}^{\mu}\hat{k}^{\nu}e^{-(\hat{\bf k}\cdot {\bf x})^{2}},
\label{diff4}
\end{eqnarray}
where we have noted $\hat{\bf k}={\bf k}/k$. In $d=3$, introducing a
spherical system of coordinates with the $z$-axis in the direction of
${\bf x}$, we obtain
\begin{eqnarray}
G^{\mu\nu}=(G_{\|}-{1\over 2}G_{\perp}){x^{\mu}x^{\nu}\over x^{2}}+{1\over 2}G_{\perp}\delta^{\mu\nu},
\label{diff5}
\end{eqnarray}
with
\begin{eqnarray}
G_{\|}={2\pi^{3/2}\over x}G(x),\quad
G_{\perp}={2\pi^{3/2}\over x}\lbrack {\rm \Phi}(x)-G(x)\rbrack,
\label{diff6}
\end{eqnarray}
where
\begin{eqnarray}
G(x)={2\over\sqrt{\pi}}{1\over x^{2}}\int_{0}^{x}t^{2}e^{-t^{2}}dt=\frac{\Phi(x)-x\Phi'(x)}{2x^{2}},\nonumber\\
\label{diff7}
\end{eqnarray}
and $\Phi(x)$ is the error function
\begin{eqnarray}
{\rm \Phi}(x)={2\over \sqrt{\pi}}\int_{0}^{x}e^{-t^{2}}dt.
\label{diff8}
\end{eqnarray}

\begin{figure}
\centering
\includegraphics[width=8cm]{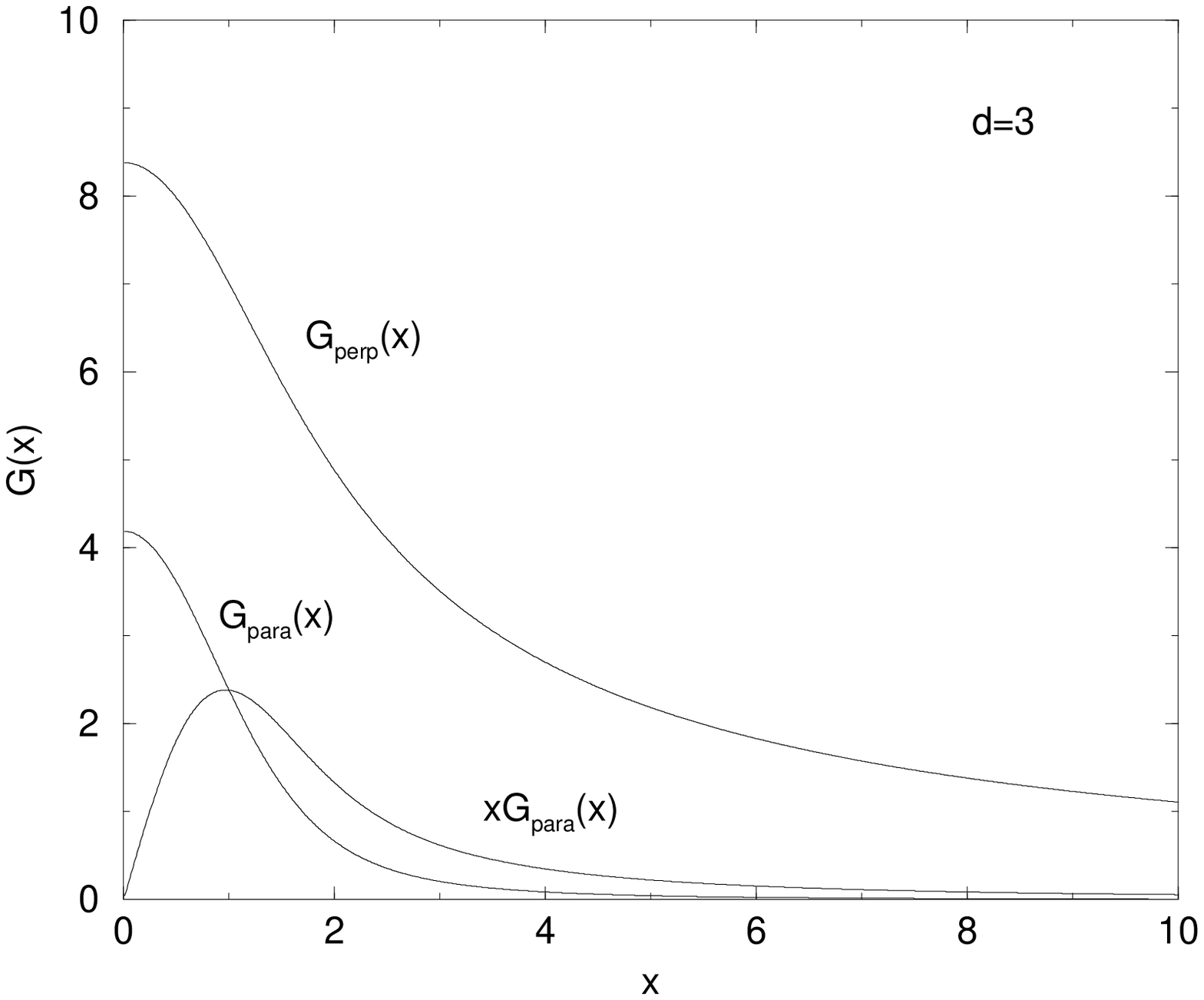}
\caption{Diffusion coefficients $G_{\|}(x)$, $G_{\perp}(x)$ and friction force $xG_{\|}(x)$ for a thermal bath in $d=3$. }
\label{dim3}
\end{figure}

\begin{figure}
\centering
\includegraphics[width=8cm]{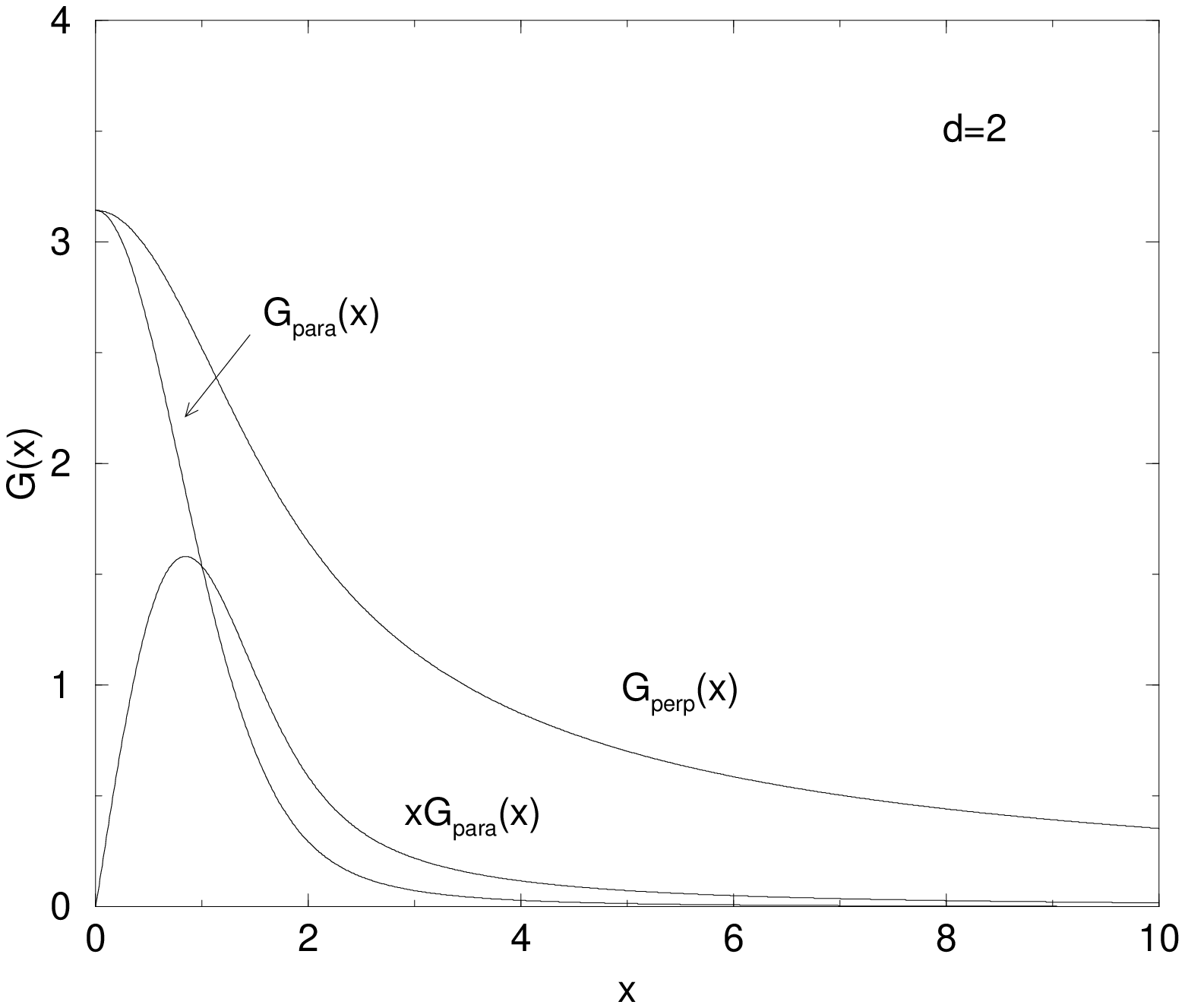}
\caption{Diffusion coefficients $G_{\|}(x)$, $G_{\perp}(x)$ and friction force $xG_{\|}(x)$ for a thermal bath in $d=2$. }
\label{dim2}
\end{figure}

\begin{figure}
\centering
\includegraphics[width=8cm]{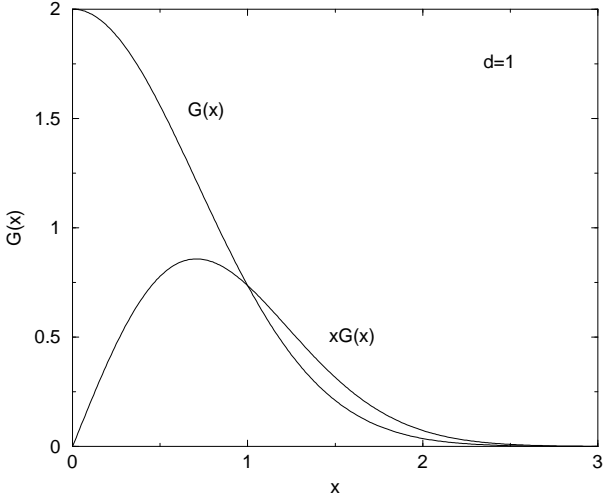}
\caption{Diffusion coefficient $G(x)$ and friction force $xG(x)$ for a thermal bath in $d=1$. }
\label{dim1}
\end{figure}

In $d=2$, using a polar system of coordinates with the $x$-axis in the direction of ${\bf x}$, we get
\begin{eqnarray}
G^{\mu\nu}=(G_{\|}-G_{\perp}){x^{\mu}x^{\nu}\over x^{2}}+G_{\perp}\delta^{\mu\nu},
\label{diff9}
\end{eqnarray}
with
\begin{eqnarray}
G_{\|}=\pi e^{-{x^{2}\over 2}}\biggl\lbrack I_{0}\biggl ({x^{2}\over 2}\biggr )- I_{1}\biggl ({x^{2}\over 2}\biggr )\biggr\rbrack,
\label{diff10}
\end{eqnarray}
\begin{eqnarray}
G_{\perp}=\pi e^{-{x^{2}\over 2}}\biggl\lbrack I_{0}\biggl ({x^{2}\over 2}\biggr )+I_{1}\biggl ({x^{2}\over 2}\biggr )\biggr\rbrack.
\label{diff11}
\end{eqnarray}

Finally, in $d=1$, we obtain
\begin{eqnarray}
G(x)=2 e^{-x^{2}}.
\label{diff12}
\end{eqnarray}
The diffusion coefficients and friction force are plotted in Figs. \ref{dim3}-\ref{dim1} for different dimensions of space.

\subsection{The relaxation time}
\label{sec_time}

We can use the above results to estimate the relaxation time of the
test particle towards the Maxwellian distribution (thermalization).
We consider the relaxation of a test particle with mass $m$ in a
thermal bath of field particles with mass $m_{f}$. If we set ${\bf
x}=\sqrt{\beta m_{f}/2}{\bf v}$, the Fokker-Planck equation
(\ref{einstein5}) can be rewritten
\begin{eqnarray}
\label{r1}
\frac{\partial P}{\partial t}=\frac{1}{t_{R}}\frac{\partial}{\partial x^{\mu}}\left\lbrack G^{\mu\nu}(x)\left (\frac{\partial P}{\partial x^{\nu}}+2\frac{m}{m_{f}} P x^{\nu}\right )\right\rbrack,
\end{eqnarray}
where $t_{R}$ is a reference time given by
\begin{eqnarray}
\label{r2}
\frac{1}{t_{R}}=\left (\frac{\pi}{8}\right )^{1/2}d^{3/2}(2\pi)^{d}\frac{\rho m_{f}}{v_{mf}^{3}}\int_{0}^{+\infty}k^{d}\hat{u}(k)^{2}dk,
\end{eqnarray}
where $v_{mf}^{2}=d/(\beta m_{f})$ is the mean-squared velocity of the
field particles.
Introducing the dimensionless function
\begin{eqnarray}
\label{r3}
\eta(k)=-(2\pi)^{d}\hat{u}(k)\beta m_{f}\rho,
\end{eqnarray} 
we can rewrite the reference time (\ref{r2}) in the form
\begin{eqnarray}
\label{r4}
\frac{1}{t_{R}}=\left (\frac{\pi}{8d}\right )^{1/2}\frac{v_{mf}}{nL^{d+1}}\frac{1}{(2\pi)^{d}}\int_{0}^{+\infty}\kappa^{d}\eta(\kappa/L)^{2}d\kappa,
\end{eqnarray}
where $L=V^{1/d}$ is the size of the system and $\kappa=k L$ is dimensionless ($n=\rho/m_{f}$ is the numerical density of the field particles). We define a typical dynamical time by
\begin{eqnarray}
\label{r5}
t_{D}=\frac{L}{v_{mf}}.
\end{eqnarray}
Then, introducing the number $N=nL^{d}$ of field particles, we can finally write the reference time in the form
\begin{eqnarray}
\label{r6}
t_{R}=C_{d} Nt_{D},
\end{eqnarray}
where 
\begin{eqnarray}
\label{r7}
C_{d}^{-1}=\left (\frac{\pi}{8d}\right )^{1/2}\frac{1}{(2\pi)^{d}}\int_{0}^{+\infty}\kappa^{d}\eta(\kappa/L)^{2}d\kappa,
\end{eqnarray}
is a dimensionless number. 

We can get an estimate of the relaxation time by the following argument. If the diffusion coefficient were constant, the typical velocity of the test particle (in one spatial direction) would increase like
\begin{eqnarray}
\label{r8}
\frac{1}{d}\langle (\Delta {\bf v})^{2}\rangle\sim 2D_{\|}t.
\end{eqnarray}
The relaxation time $t_{r}$ is the typical time at which the typical velocity 
of the test particle has reached its equilibrium value $\langle v^{2}\rangle(+\infty)=d/(m\beta)=(m_{f}/m)v_{mf}^{2}$ so that $\langle (\Delta {\bf v})^{2}\rangle(t_{r})=\langle v^{2}\rangle(+\infty)$. Since $D_{\|}$ depends on $v$, the description of the diffusion is more complex. However, the formula 
\begin{eqnarray}
\label{r9}
t_{r}= \frac{1}{d} \frac{m_{f}}{m}\frac{v_{mf}^{2}}{2D_{\|}(v_{mf})},
\end{eqnarray}
resulting from the previous arguments with $D_{\|}=D_{\|}(v_{mf})$ 
should provide a good estimate of the relaxation time. 
Using Eq. (\ref{diff3}) and comparing with Eq. (\ref{r4}) we obtain 
\begin{eqnarray}
\label{r10}
t_{r}=K_{d} \frac{m_{f}}{m}t_{R},
\end{eqnarray}
where $K_{d}=1/\lbrack 4G_{\|}(\sqrt{d/2})\rbrack$. We find $K_{3}=0.13587547...$, $K_{2}=0.16286327...$ and $K_{1}=0.20609016...$. We can also estimate the relaxation time by $t_{r}'=\xi^{-1}$ where $\xi$ is the friction coefficient. Using the Einstein relation $\xi=D_{\|}\beta m$ with  $D_{\|}=D_{\|}(v_{mf})$ we find that 
\begin{eqnarray}
\label{r11}
t'_{r}=2t_{r}.
\end{eqnarray}

Combining the previous results we find that the relaxation time scales as
\begin{eqnarray}
\label{r12}
t_{r}\sim  N \frac{m_{f}}{m} t_{D}.
\end{eqnarray}
We expect that this result also yields a good estimate of the
relaxation time of a test particle towards the stationary distribution
(\ref{iso8}) in the case of a non-thermal bath. In that case, we must
justify that the DF of the bath does not change on that interval (see
below). We come therefore to the following conclusions. For equal mass
particles, the relaxation time of a test particle in a bath scales as
$Nt_{D}$. For $d=2,3$, the relaxation time of the system as a whole
also scales as $Nt_{D}$ (see Sec. \ref{sec_landau}). Therefore, a
non-thermal bath will change on this timescale. Only the Maxwellian
distribution is stationary on the timescale $Nt_{D}$. Thus, for
$m=m_{f}$, the test particle approach can be developed only for a
thermal bath.  Now, consider the relaxation of a test particle with
mass $m$ in a bath of field particles with mass $m_{f}$. In that case,
the relaxation of the test particle is changed by a factor
$m_{f}/m$. If $m\gg m_{f}$ we find that the relaxation of the test
particle ($\sim (m_{f}/m)Nt_{D}$) towards Eq.  (\ref{iso8}) is much
faster than the relaxation of the system of field particles as a whole
($\sim Nt_{D})$ towards the Maxwellian $f^{eq}$. Therefore, in that
limit, when we focus on the evolution of a test particle, it is
possible to consider that the distribution of the field particles
$f({\bf v},t)$ is ``frozen'' even if it does not correspond to
statistical equilibrium. This is because $f({\bf v},t)$ evolves much
slower than $P({\bf v},t)$. This remark justifies to consider
equations of the form (\ref{fp1}) with $f\neq f^{eq}$. In
astrophysics, the case $m\gg m_{f}$ could be relevant to describe the
stochastic evolution of a black hole at the center of a galaxy or the
dynamics of a globular cluster ($m\sim 10^{6}m_{f}$) passing through a
galaxy. Note that in $d=1$ the relaxation of a test particle in a bath
is given by Eq. (\ref{r12}) while the relaxation of the field
particles as a whole towards statistical equilibrium is {\it larger}
than $Nt_{D}$ since the Lenard-Balescu collision term cancels out in
$d=1$. Therefore, in that case it is justified to consider the
relaxation of a test particle in a bath with {\it any} distribution
$f(v)$ (stable with respect to the Vlasov equation) even if $m=m_{f}$
(see Sec. \ref{sec_d1}).

\subsection{Examples}
\label{sec_examples}

Let us apply these results to particular systems considered in
\cite{hb1,hb2}.  For the gravitational interaction in $d=3$, the Fourier
transform of the potential is
\begin{eqnarray}
\label{ex1}
(2\pi)^{3}\hat{u}(k)=-\frac{4\pi G}{k^{2}}.
\end{eqnarray}
The reference time (\ref{r2}) can be written
\begin{eqnarray}
\label{ex2}
{t_{R}}=0.482\frac{v_{mf}^{3}}{nm_{f}^{2}G^{2}\ln\Lambda},
\end{eqnarray}
where $\ln\Lambda$ is the Coulombian factor \cite{bt}. Using
Eqs. (\ref{r3}) and (\ref{ex1}), we get
\begin{eqnarray}
\label{ex3}
\eta(k)=\frac{k_{J}^{2}}{k^{2}}, 
\end{eqnarray}
where $k_{J}=(4\pi G\beta\rho m_{f})^{1/2}$ is the Jeans wavenumber. Then,
Eqs. (\ref{r6})-(\ref{r7}) lead to
\begin{eqnarray}
\label{ex4}
t_{R}=\frac{11.3}{\eta^{2}}\frac{N}{\ln\Lambda}t_{D},
\end{eqnarray}
where we have defined $\eta=\beta GNm_{f}^{2}/R$ with $L^{3}=V=(4/3)\pi R^{3}$ and we recall that $\ln\Lambda\sim \ln N$.

For the HMF model in $d=1$, we have
\begin{eqnarray}
\label{ex5}
\hat{u}_{n}=-\frac{k}{4\pi}(\delta_{n,1}+\delta_{n,-1}).
\end{eqnarray}
The reference time  (\ref{r2}) can be written
\begin{eqnarray}
\label{ex6}
t_{R}=40.1\frac{v_{mf}^{3}}{nm_{f}^{2}k^{2}},
\end{eqnarray}
where we have replaced $\int_{0}^{+\infty} dk$ by
$\sum_{n=0}^{+\infty}$ in the discrete case.  Using Eqs. (\ref{r3}) and
(\ref{ex5}), we get
\begin{eqnarray}
\label{ex7}
\eta_{n}=\eta (\delta_{n,1}+\delta_{n,-1}), \qquad \eta=\frac{\beta Nm_{f}^{2}k}{4\pi}.
\end{eqnarray}
Then, we obtain 
\begin{eqnarray}
\label{ex8}
t_{R}=\frac{0.254}{\eta^{2}}Nt_{D}.
\end{eqnarray}
For the HMF model, the relaxation towards the Maxwellian is not
exponential due to the rapid decay of the diffusion coefficient
\cite{bd,cl} so that this reference time only gives a timescale of
relaxation.

\section{The case $d=3$}
\label{sec_d3}

\subsection{Rosenbluth potentials}
\label{sec_rosenbluth}

In $d=3$, it is possible to obtain a simple expression of the
coefficients of diffusion and friction, expressed in terms of
elementary integrals, for {\it any} isotropic distribution of the bath
(not only the Maxwellian). To that purpose, it is useful to introduce
the so-called Rosenbluth potentials
\cite{rosen,bt,genlandau}. Introducing a spherical system of
coordinates with the $z$-axis in the direction of ${\bf u}$, we find
after elementary calculations that Eq. (\ref{fp3}) can be written
\begin{eqnarray}
\label{rosen1}
K^{\mu\nu}=A_{3}\frac{\delta^{\mu\nu}u^{2}-u^{\mu}u^{\nu}}{u^{3}}
\end{eqnarray}
where
\begin{eqnarray}
\label{rosen2}
A_{3}=8\pi^{5}\int_{0}^{+\infty}k^{3}\hat{u}(k)^{2}dk.
\end{eqnarray}
In particular, for the gravitational potential we find that $A_{3}=2\pi G^{2}\ln\Lambda$ where $\ln\Lambda=\int_{0}^{+\infty}dk/k$ is the Coulombian factor which has to be regularized with appropriate cut-offs (see the Introduction). Using Eq. (\ref{rosen1}), the Landau equation (\ref{landau6}) can be written
\begin{eqnarray}
\label{rosen3}
\frac{\partial f}{\partial t}=A\frac{\partial}{\partial v^{\mu}}\int \frac{\delta^{\mu\nu}u^{2}-u^{\mu}u^{\nu}}{u^{3}}\left (f'\frac{\partial f}{\partial v^{\nu}}-f\frac{\partial f'}{\partial v'^{\nu}}\right )d{\bf v}'\nonumber\\
\end{eqnarray} 
with $A=2\pi G^{2}m\ln\Lambda$. Equation (\ref{rosen3}) is the
original form of the Landau equation (we have used the notations of
astrophysics but the connection with the notations of plasma physics
is obtained by making the substitution $Gm^{2}\leftrightarrow
-e^{2}$). It applies to weakly coupled systems. We note that the
potential of interaction appears only in the constant $A$ which merely
determines the timescale of relaxation. The structure of the Landau equation
is independent on the potential. Using the identity
\begin{eqnarray}
\label{rosen4}
\frac{\partial^{2}u}{\partial v^{\mu}\partial v^{\nu}}=\frac{\delta^{\mu\nu}u^{2}-u^{\mu}u^{\nu}}{u^{3}},
\end{eqnarray}
we can write the diffusion tensor (\ref{fp4}) in the form
\begin{eqnarray}
\label{rosen5}
D^{\mu\nu}=A_{3}\sum_{j} m_{j}\frac{\partial^{2}g_{j}}{\partial v^{\mu}\partial v^{\nu}} ({\bf v}),
\end{eqnarray}
where 
\begin{eqnarray}
\label{rosen6}
g({\bf v})=\int f({\bf v}')|{\bf v}-{\bf v}'|d{\bf v}'.
\end{eqnarray}
On the other hand, integrating by parts and using the fact that $K^{\mu\nu}$ depends only on the relative velocity ${\bf u}={\bf v}-{\bf v}'$, the friction term (\ref{fp5}) can be written
\begin{eqnarray}
\label{rosen7}
\eta^{\mu}=-m\sum_{j}\int \frac{\partial K^{\mu\nu}}{\partial v^{\nu}} f'_{j}d{\bf v}'.
\end{eqnarray}
Using the identity
\begin{eqnarray}
\label{rosen8}
\frac{\partial K^{\mu\nu}}{\partial v^{\nu}}=-2A_{3}\frac{u^{\mu}}{u^{3}}=2A_{3}  \frac{\partial}{\partial v^{\mu}}\left (\frac{1}{u}\right),
\end{eqnarray}
we can rewrite the friction term in the form
\begin{eqnarray}
\label{rosen9}
\eta^{\mu}=-2 A_{3}m\sum_{j} \frac{\partial h_{j}}{\partial v^{\mu}} ({\bf v}),
\end{eqnarray}
where 
\begin{eqnarray}
\label{rosen10}
h({\bf v})=\int \frac{f({\bf v}')}{|{\bf v}-{\bf v}'|}d{\bf v}'.
\end{eqnarray}
The functions $g({\bf v})$ and $h({\bf v})$ are called the Rosenbluth
potentials. Using Eqs. (\ref{rosen5}) and (\ref{rosen9}), we can
readily check that the first and second moments of the velocity
increments given by Eqs. (\ref{fp8}) and (\ref{fp11}) return the
expressions obtained by Rosenbluth {\it et al.}
\cite{rosen} and that the kinetic equation derived in \cite{rosen} is
equivalent to Eq. (\ref{rosen3}), although written in a different
form. Therefore, we have recovered the kinetic equation of stellar
dynamics directly from the Landau equation. This shows that the Landau
equation is equivalent to the kinetic equation derived by
Chandrasekhar \cite{chandra} and Rosenbluth {\it et al.} \cite{rosen}
starting directly from the Fokker-Planck equation and evaluating the
first and second moments of the velocity increments by considering a
succession of binary encounters. This was expected because the Landau
equation carries the same type of assumptions. However, it is
surprising that the relation to the Landau equation was not mentioned
in \cite{chandra,rosen,bt}. In particular, these authors write the
Fokker-Planck equation in the form (\ref{fp7}) with the diffusion
coefficient in the second derivative
($\partial_{\mu}\partial_{\nu}D^{\mu\nu}$) while a more symmetric form
is the Landau equation (\ref{rosen3}) where the diffusion coefficient
is inserted between the first derivatives
($\partial_{\mu}D^{\mu\nu}\partial_{\nu}$), see Eq. (\ref{fp6}). It
is interesting to note, for historical reasons, that this symmetric
form (from which we immediately deduce all the conservation laws of the
system and the $H$-theorem
\cite{balescubook}) has escaped to the study of stellar dynamicists \cite{chandra,rosen,bt} while the Landau
equation was known long before in plasma physics.

\subsection{Diffusion and friction}
\label{sec_difffric}

We shall see that the Rosenbluth potentials can be easily calculated
in $d=3$ when the field particles have an isotropic velocity
distribution. The results of the previous section remain valid in
$d=2$ provided that we replace $A_{3}$ by $A_{2}$ defined in
Eq. (\ref{gen14}). However, the calculation of the Rosenbluth
potentials is apparently more complicated in $d=2$ than in $d=3$ and
this is why we shall use another method in $d=2$ to obtain the
diffusion and friction coefficients (see Sec. \ref{sec_genexp}).

If the field particles have an isotropic velocity distribution, the
Rosenbluth potentials in $d=3$ take the particularly simple form
\cite{rosen,bt,genlandau}:
\begin{eqnarray}
\label{di1}
h(v)=4\pi\left\lbrack \frac{1}{v}\int_{0}^{v}f(v_{1})v_{1}^{2}dv_{1}+\int_{v}^{+\infty} f(v_{1})v_{1}dv_{1}\right \rbrack,
\end{eqnarray}
\begin{eqnarray}
\label{di2}
g(v)=\frac{4\pi v}{3}\biggl\lbrack \int_{0}^{v}\left (3 v_{1}^{2}+\frac{v_{1}^{4}}{v^{2}}\right ) f(v_{1})dv_{1}\nonumber\\
+\int_{v}^{+\infty}\left (\frac{3v_{1}^{3}}{v}+v v_{1}\right ) f(v_{1})dv_{1}\biggr \rbrack.
\end{eqnarray}
Furthermore, when $g=g(v)$ the diffusion tensor (\ref{rosen5}) can be
put in the form of Eq. (\ref{iso1}) with
\begin{eqnarray}
\label{di3}
D_{\|}=A_{3}\sum_{j} m_{j}\frac{d^{2}g_{j}}{dv^{2}},
\end{eqnarray}
\begin{eqnarray}
\label{di4}
D_{\perp}=2A_{3}\sum_{j} m_{j}\frac{1}{v}\frac{dg_{j}}{dv}.
\end{eqnarray}
Using Eq. (\ref{di2}) we obtain
\begin{eqnarray}
\label{di5}
D_{\|}=\frac{8\pi}{3}A_{3}\sum_{j} m_{j}\frac{1}{v}\biggl\lbrack \int_{0}^{v}\frac{v_{1}^{4}}{v^{2}} f_{j}(v_{1})dv_{1}\nonumber\\
+v\int_{v}^{+\infty}v_{1}f_{j}(v_{1})dv_{1}\biggr \rbrack,
\end{eqnarray}
\begin{eqnarray}
\label{di6}
D_{\perp}=\frac{8\pi}{3}A_{3}\sum_{j}m_{j}\frac{1}{v}\biggl\lbrack \int_{0}^{v}\left (3 v_{1}^{2}-\frac{v_{1}^{4}}{v^{2}}\right ) f_{j}(v_{1})dv_{1}\nonumber\\
+2v\int_{v}^{+\infty}v_{1}f_{j}(v_{1})dv_{1}\biggr \rbrack.
\end{eqnarray}
On the other hand, when $h=h(v)$ the friction term (\ref{rosen9}) can
be written
\begin{eqnarray}
\label{di7}
{\mb \eta}=-2A_{3}m\sum_{j}\frac{1}{v}\frac{dh_{j}}{dv}{\bf v}.
\end{eqnarray}
Using  Eq. (\ref{di1}) we obtain
\begin{eqnarray}
\label{di8}
{\mb \eta}=8\pi A_{3}m\frac{\bf v}{v^{3}}\sum_{j} \int_{0}^{v}f_{j}(v_{1})v_{1}^{2}dv_{1}.
\end{eqnarray}

We note that these expressions are valid for {\it any} isotropic
distribution of the field particles. If we substitute Eqs. (\ref{di5})
and (\ref{di8}) into Eq. (\ref{iso4}), we get the Fokker-Planck
equation describing the evolution of a test particle in a bath with
prescribed distribution $f_{j}(v)$. Alternatively, if we come back to
the original Landau kinetic equation (\ref{landau6}), assume an
isotropic velocity distribution and substitute the general expressions
(\ref{di5}) and (\ref{di8}) with now $f_{j}=f_{j}(v,t)$ we obtain the
integro-differential equation
\begin{eqnarray}
\label{di9}
\frac{\partial f_{i}}{\partial t}=8\pi A_{3}\sum_{j}\frac{1}{v^{2}}\frac{\partial}{\partial v}\biggl\lbrack \frac{m_{j}}{3}\frac{\partial f_{i}}{\partial v}\biggl ( \frac{1}{v}\int_{0}^{v}{v_{1}^{4}} f_{j}(v_{1},t)dv_{1}\nonumber\\
+v^{2}\int_{v}^{+\infty}v_{1}f_{j}(v_{1},t)dv_{1}\biggr )+m_{i}f_{i}\int_{0}^{v}f_{j}(v_{1},t)v_{1}^{2}dv_{1}\biggr\rbrack,\nonumber\\
\end{eqnarray}
describing the evolution of the system as a whole.

\subsection{Water-bag distribution}
\label{sec_waterbag}

\begin{figure}
\centering
\includegraphics[width=8cm]{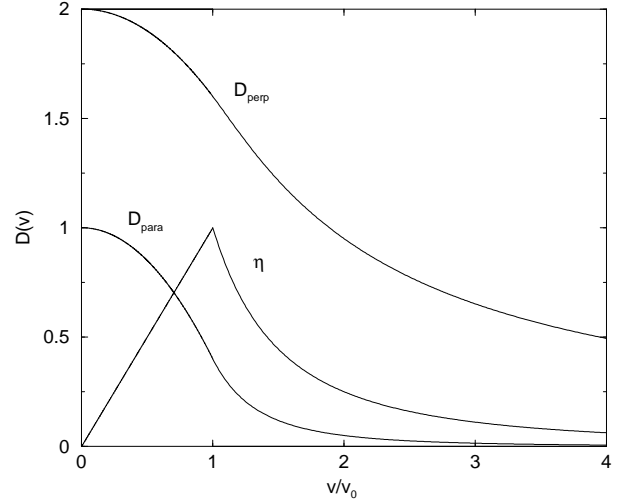}
\caption{Diffusion coefficients $D_{\|}(v)/D_{*}$, $D_{\perp}(v)/D_{*}$ and friction force $\eta(v)/\eta_{*}$ as a function of $v/v_{0}$ for a water-bag distribution of the bath in $d=3$. The normalization factors are $D_{*}=\frac{4\pi}{3}A_{3}m_{f}\eta_{0}v_{0}^{2}$ and $\eta_{*}=\frac{8\pi}{3}A_{3}m\eta_{0}v_{0}$.}
\label{diffwater3}
\end{figure}

As an illustration of the previous formalism, we shall compute the
diffusion coefficient and friction force of a test particle with mass
$m$ when the distribution function of the bath (composed of particles
with mass $m_{f}$) is a step function: $f({\bf v})=\eta_{0}$ for $v\le
v_{0}$ and $f({\bf v})=0$ for $v>v_{0}$ with $\eta_{0}=3\rho/(4\pi
v_{0}^{3})$ ($\rho$ is the mass density of the field particles). Using
Eqs. (\ref{di5}), (\ref{di6}) and (\ref{di8}) we get for $v\le v_{0}$:
\begin{eqnarray}
\label{wb1}
D_{\|}=\frac{4\pi}{3}A_{3} m_{f} \eta_{0} \left (v_{0}^{2}-\frac{3}{5}v^{2}\right ),
\end{eqnarray}
\begin{eqnarray}
\label{wb2}
D_{\perp}=\frac{8\pi}{3}A_{3} m_{f} \eta_{0} \left (v_{0}^{2}-\frac{v^{2}}{5}\right ),
\end{eqnarray}
\begin{eqnarray}
\label{wb3}
\eta=\frac{8\pi}{3}A_{3}m \eta_{0} v.
\end{eqnarray}
and for $v>v_{0}$:
\begin{eqnarray}
\label{wb4}
D_{\|}=\frac{8\pi}{15}A_{3} m_{f} \eta_{0} v_{0}^{5}\frac{1}{v^{3}},
\end{eqnarray}
\begin{eqnarray}
\label{wb5}
D_{\perp}=\frac{8\pi}{3}A_{3} m_{f} \eta_{0} \frac{1}{v}\left (v_{0}^{3}-\frac{v_{0}^{5}}{5v^{2}}\right ),
\end{eqnarray}
\begin{eqnarray}
\label{wb6}
\eta=\frac{8\pi}{3}A_{3}m \eta_{0} \frac{v_{0}^{3}}{v^{2}}.
\end{eqnarray}
These quantities are plotted in Fig. \ref{diffwater3}. We note that
the friction term and the diffusion coefficient in the direction
parallel to the direction of the test particle are related to each
other by
\begin{eqnarray}
\label{wb7}
\eta=\frac{2mD_{\|}}{m_{f}(v_{0}^{2}-\frac{3}{5}v^{2})}v,\qquad (v\le v_{0}),
\end{eqnarray}
\begin{eqnarray}
\label{wb8}
\eta=\frac{5mD_{\|}}{m_{f}v_{0}^{2}}v,\qquad (v>v_{0}).
\end{eqnarray}
These expresssions can be compared to the Einstein relation
(\ref{einstein7}). We note that the role of the temperature is played here by
$m_{f} v_{0}^{2}$. 

Substituting these relations in Eq. (\ref{iso7}) we get a
Fokker-Planck equation (\ref{iso6}) with an effective potential
\begin{eqnarray}
\label{wb9}
U(v)=\frac{5}{3}\frac{m}{m_{f}}\left\lbrack \frac{3}{2}-\ln\left (\frac{5}{2}-\frac{3}{2}\frac{v^{2}}{v_{0}^{2}}\right )\right\rbrack,\quad (v\le v_{0}),
\end{eqnarray}
\begin{eqnarray}
\label{wb10}
U(v)=\frac{5}{2}\frac{m}{m_{f}}\frac{v^{2}}{v_{0}^{2}},\quad (v> v_{0}).
\end{eqnarray}
The stationary solution of the Fokker-Planck equation is
\begin{eqnarray}
\label{wb11}
P^{eq}(v)=Ae^{-\frac{5}{2}\frac{m}{m_{f}}}\left (\frac{5}{2}-\frac{3}{2}\frac{v^{2}}{v_{0}^{2}}\right )^{\frac{5m}{3m_{f}}}\quad (v\le  v_{0})
\end{eqnarray}
\begin{eqnarray}
\label{wb12}
P^{eq}(v)=A e^{-\frac{5}{2}\frac{m}{m_{f}}\left (\frac{v}{v_{0}}\right )^{2}}\quad (v> v_{0})
\end{eqnarray}
where $A$ is a normalization constant. We explicitly check on this
example that the stationary velocity distribution of the test particle
is different from that of the bath. As explained at the end of
Sec. \ref{sec_time}, this is just a quasi-stationary distribution
valid on a timescale $(m_{f}/m)Nt_{D}\ll Nt_{D}$ for $m\gg m_{f}$
because on longer timescales ($\sim N t_{D}$) the bath distribution
will change.

We can also compute the diffusion coefficient and the friction force
from Eqs. (\ref{di5}), (\ref{di6}) and (\ref{di8}) for a Maxwellian
velocity distribution of the field particles. In that case, we
directly check that the Einstein relation (\ref{einstein7}) holds and
that the diffusion coefficients are given by Eqs. (\ref{diff6}) and
(\ref{diff7}). For comparison with the water-bag model, we give below the
asymptotic expressions of the diffusion coefficient and friction force
for a thermal bath. For $v\rightarrow +\infty$, we have
\begin{eqnarray}
D_{\|}=2 A_{3}\frac{\rho}{\beta}\frac{1}{v^{3}},
\label{xc1}
\end{eqnarray}
\begin{eqnarray}
D_{\perp}=2 A_{3}\rho m_{f} \frac{1}{v},
\label{xc2}
\end{eqnarray}
\begin{eqnarray}
\eta=2 A_{3}\rho m \frac{1}{v^{2}}.
\label{xc3}
\end{eqnarray}
For $v\rightarrow 0$, we get
\begin{eqnarray}
D_{\|}=\frac{4}{3}A_{3}\left (\frac{\beta}{2\pi}\right )^{1/2}\rho m_{f}^{3/2}\left (1-\frac{3}{10}\beta m_{f}v^{2}+...\right ),\nonumber\\
\label{xc4}
\end{eqnarray}
\begin{eqnarray}
D_{\perp}=\frac{8}{3}A_{3}\left (\frac{\beta}{2\pi}\right )^{1/2}\rho m_{f}^{3/2}\left (1-\frac{1}{10}\beta m_{f}v^{2}+...\right ),\nonumber\\
\label{xc5}
\end{eqnarray}
\begin{eqnarray}
\eta=\frac{4}{3}A_{3}\left (\frac{\beta}{2\pi}\right )^{1/2}\rho m_{f}^{3/2}\beta m v.
\label{xc6}
\end{eqnarray}
Interestingly, these are the same asymptotic behaviors (same scaling)
as in the water-bag model. For recent studies concerning the
motion of a ``test particle'' in a thermal bath in relation
with space plasmas, and for some numerical solutions of the
corresponding Fokker-Planck equation see, e.g., Shizgal
\cite{shizgal} and references therein.

\section{The case  $d=2$}
\label{sec_d2}

\subsection{General expressions}
\label{sec_genexp}

We shall provide here general expressions of the diffusion coefficient
and friction force in $d=2$ for any isotropic velocity distribution of
the field particles. We shall use a method different from that exposed
in Sec. \ref{sec_d3} for $d=3$. The following method also works in
$d=3$ as shown in Appendix \ref{sec_alternative}.  

Using
Eq. (\ref{fp3}), the diffusion coefficient (\ref{fp4}) can be written
\begin{eqnarray}
D^{\mu\nu}=\pi (2\pi)^{d}\sum_{j}m_{j}\int d{\bf k}d{\bf v}' k^{\mu}k^{\nu}\hat{u}(k)^{2}\delta({\bf k}\cdot{\bf u})f_{j}({\bf v}').\nonumber\\
\label{gen1}
\end{eqnarray} 
Inserting the identity (\ref{diff1}) in Eq. (\ref{gen1}), we get
\begin{eqnarray}
D^{\mu\nu}=(2\pi)^{2d}\sum_{j}m_{j}\int_{0}^{+\infty}dt\int d{\bf k} k^{\mu}k^{\nu}\hat{u}(k)^{2}e^{i{\bf k}\cdot {\bf v}t}\hat{f}_{j}({\bf k}t),\nonumber\\
\label{gen2}
\end{eqnarray} 
where we have introduced the Fourier transform
\begin{eqnarray}
\hat{f}({\bf k})=\int f({\bf x})e^{-i{\bf k}\cdot{\bf x}}\frac{d{\bf x}}{(2\pi)^{d}}.
\label{gen3}
\end{eqnarray} 
For an isotropic velocity distribution, $\hat{f}({\bf k})=\Phi(k)$
depends only on the modulus of ${\bf k}$ (see the explicit expression
below). Setting $\tau=kt$ we finally obtain
\begin{eqnarray}
D^{\mu\nu}=(2\pi)^{2d}\sum_{j}m_{j}\int_{0}^{+\infty}d\tau\int d{\bf k} \frac{k^{\mu}k^{\nu}}{k}\hat{u}(k)^{2}e^{i\hat{\bf k}\cdot {\bf v}\tau}{\Phi}_{j}(\tau)\nonumber\\
\label{gen4}
\end{eqnarray} 
where $\hat{\bf k}={\bf k}/k$. This expression is valid in $d$ dimensions. We now specialize on the case $d=2$. Introducing polar coordinates with the $x$-axis in the direction of ${\bf v}$ and using the identities
\begin{eqnarray}
\int_{0}^{2\pi}e^{ix\cos\theta}d\theta=2\pi J_{0}(x),
\label{gen5}
\end{eqnarray} 
\begin{eqnarray}
\int_{0}^{2\pi}e^{ix\cos\theta}\sin^{2}\theta d\theta=2\pi \frac{J_{1}(x)}{x},
\label{gen6}
\end{eqnarray}
we find that $D^{\mu\nu}$ is given by Eq. (\ref{iso1}) with
\begin{eqnarray}
D_{\|}=(2\pi)^{5}\sum_{j}m_{j}\frac{1}{v}\int_{0}^{+\infty}k^{2}\hat{u}(k)^{2}dk\nonumber\\
\times\int_{0}^{+\infty}dx\left\lbrack J_{0}(x)-\frac{J_{1}(x)}{x}\right\rbrack\Phi_{j}\left (\frac{x}{v}\right),
\label{gen7}
\end{eqnarray}
\begin{eqnarray}
D_{\perp}=(2\pi)^{5}\sum_{j}m_{j}\frac{1}{v}\int_{0}^{+\infty}k^{2}\hat{u}(k)^{2}dk\nonumber\\
\times\int_{0}^{+\infty}dx \frac{J_{1}(x)}{x}\Phi_{j}\left (\frac{x}{v}\right).
\label{gen8}
\end{eqnarray}
Now, in $d=2$, 
\begin{eqnarray}
\Phi(k)=\frac{1}{2\pi}\int_{0}^{+\infty}f(v_{1})J_{0}(k v_{1})v_{1}dv_{1}.
\label{gen9}
\end{eqnarray}
Substituting this expression in Eqs. (\ref{gen7}) and (\ref{gen8}) and
introducing the notations
\begin{eqnarray}
I_{\|}(\lambda)=\int_{0}^{+\infty} \left\lbrack J_{0}(x)-\frac{J_{1}(x)}{x}\right\rbrack J_{0}(\lambda x)dx,
\label{gen10}
\end{eqnarray}
\begin{eqnarray}
I_{\perp}(\lambda)=\int_{0}^{+\infty}\frac{J_{1}(x)}{x} J_{0}(\lambda x)dx, 
\label{gen11}
\end{eqnarray}
we obtain
\begin{eqnarray}
D_{\|}=2\pi A_{2}\sum_{j}m_{j}\frac{1}{v}\int_{0}^{+\infty}  f_{j}(v_{1})I_{\|}\left (\frac{v_{1}}{v}\right )v_{1}dv_{1}, 
\label{gen12}
\end{eqnarray}
\begin{eqnarray}
D_{\perp}=2\pi A_{2}\sum_{j}m_{j}\frac{1}{v}\int_{0}^{+\infty}  f_{j}(v_{1})I_{\perp}\left (\frac{v_{1}}{v}\right )v_{1}dv_{1}, 
\label{gen13}
\end{eqnarray}
where
\begin{eqnarray}
\label{gen14}
A_{2}=8\pi^{3}\int_{0}^{+\infty}k^{2}\hat{u}(k)^{2}dk.
\end{eqnarray}
Finally, the friction is given by Eq. (\ref{iso3}). Note that these
expressions are valid for an arbitrary isotropic velocity distribution of the
field particles. They remain true if the distribution function depends
on time. Therefore, by substituting these results in the Landau
equation (\ref{landau6}) we obtain a self-consistent kinetic equation for an
isotropic distribution function $f(v,t)$ which is the counterpart in $d=2$
of Eq. (\ref{di9}) in $d=3$.  However, because it involves complicated
functions (\ref{gen10}) and (\ref{gen11}), its expression is less explicit.

\subsection{Water-bag distribution}
\label{sec_wb}

\begin{figure}
\centering
\includegraphics[width=8cm]{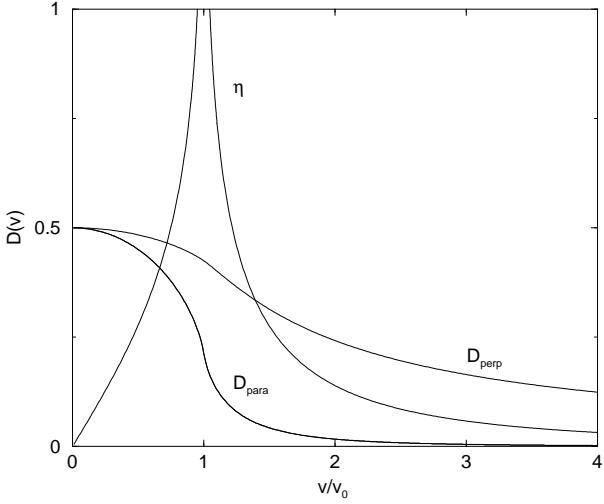}
\caption{Diffusion coefficients $D_{\|}(v)/D_{*}$, $D_{\perp}(v)/D_{*}$ and friction force $\eta(v)/\eta_{*}$ as a function of $v/v_{0}$ for a water bag distribution of the bath in $d=2$. The normalization factors are $D_{*}=2\pi A_{2}m_{f}\eta_{0}v_{0}$ and $\eta_{*}=(m/m_{f})(D_{*}/v_{0})$.}
\label{diffwater2}
\end{figure}

The previous relations are quite general. Let us now consider for
illustration the case where the distribution function of the field
particles (with mass $m_{f}$) is a step function: $f(v)=\eta_{0}$ for
$v\le v_{0}$ and $f(v)=0$ for $v>v_{0}$ where $\eta_{0}=\rho/(\pi
v_{0}^{2})$. Using the identity
\begin{eqnarray}
xJ_{0}(x)=\lbrack xJ_{1}(x)\rbrack',
\label{wbb1}
\end{eqnarray}
the Fourier transform (\ref{gen9}) of the distribution function is 
\begin{eqnarray}
\Phi(\xi)=\frac{\eta_{0}v_{0}}{2\pi\xi}J_{1}(\xi v_{0}).
\label{wbb2}
\end{eqnarray}
Inserting this expression in Eqs. (\ref{gen7}) and (\ref{gen8}), we obtain
\begin{eqnarray}
D_{\|}=2\pi A_{2} m_{f}\eta_{0}v_{0} R_{\|}\left (\frac{v_{0}}{v}\right),
\label{wbb3}
\end{eqnarray}
\begin{eqnarray}
D_{\perp}=2\pi A_{2} m_{f}\eta_{0}v_{0} R_{\perp}\left (\frac{v_{0}}{v}\right),
\label{wbb4}
\end{eqnarray}
where we have introduced the functions
\begin{eqnarray}
R_{\|}(\lambda)=\int_{0}^{+\infty}dx\left\lbrack J_{0}(x)-\frac{J_{1}(x)}{x}\right\rbrack \frac{1}{x}J_{1}(\lambda x),
\label{wbb5}
\end{eqnarray}
\begin{eqnarray}
R_{\perp}(\lambda)=\int_{0}^{+\infty}dx \frac{J_{1}(x)}{x^2}J_{1}(\lambda x).
\label{wbb6}
\end{eqnarray}
Using Eq. (\ref{iso3}), the friction term  is given by 
\begin{eqnarray}
\label{wbb7}
\eta=-\frac{m}{m_{f}}\left\lbrack \frac{dD_{\|}}{dv}+\frac{1}{v}\left (D_{\|}-{D_{\perp}}\right )\right\rbrack.\nonumber\\
\end{eqnarray}
The integrals (\ref{wbb5}) and (\ref{wbb6}) can be expressed in terms of hypergeometric functions. Their asymptotic behaviors are
derived in Appendix \ref{sec_asymptotic}. Using these results, we
obtain the asymptotic behaviors of the diffusion coefficients and
friction force. For $v\rightarrow +\infty$, we get
\begin{eqnarray}
D_{\|}=\frac{1}{4}\pi A_{2} m_{f}\eta_{0}v_{0}^{4} \frac{1}{v^{3}},
\label{wbb8}
\end{eqnarray}
\begin{eqnarray}
D_{\perp}=\pi A_{2} m_{f}\eta_{0}v_{0}^{2}\frac{1}{v},
\label{wbb9}
\end{eqnarray}
\begin{eqnarray}
\eta={\pi} A_{2} m\eta_{0} v_{0}^{2}\frac{1}{v^{2}}.
\label{wbb10}
\end{eqnarray}
We note that the leading term $\sim v^{-2}$ of the friction $\eta$ is
due to the diffusion in the direction perpendicular to the velocity of
the test particle, i.e. the term $D_{\perp}$ in Eq. (\ref{wbb7}). The
contribution of the diffusion coefficient in the direction parallel to
the velocity of the test particle decreases more rapidly like
$v^{-4}$. For $v\rightarrow 0$, we get
\begin{eqnarray}
D_{\|}=\pi A_{2} m_{f}\eta_{0}v_{0}\left\lbrack 1-\frac{3}{8}\left (\frac{v}{v_{0}}\right )^{2}+...\right\rbrack, 
\label{wbb11}
\end{eqnarray}
\begin{eqnarray}
D_{\perp}=\pi A_{2} m_{f}\eta_{0}v_{0}\left\lbrack 1-\frac{1}{8}\left (\frac{v}{v_{0}}\right )^{2}+...\right\rbrack, 
\label{wbb12}
\end{eqnarray}
\begin{eqnarray}
\eta=\pi m A_{2} \eta_{0} \frac{v}{v_{0}}.
\label{wbb13}
\end{eqnarray}
On the other hand, using Eqs. (\ref{wbb7}) and (\ref{afin9}), we find that the
friction coefficient diverges for $v\rightarrow v_{0}$ like
\begin{eqnarray}
\eta=2 A_{2}m \eta_{0} (2-2\gamma-\ln 2-2\psi(3/2)-\ln |1-v/v_{0}|).\nonumber\\
\label{divfric1}
\end{eqnarray}
The diffusion coefficients and friction force for a water-bag
distribution of the bath are plotted in Fig. \ref{diffwater2}.

We can also use the previous formalism in the case where the distribution of the bath is Maxwellian (thermal bath), in which case
\begin{eqnarray}
\Phi_{i}(\xi)=\frac{\rho_{i}}{(2\pi)^{2}}e^{-\frac{\xi^{2}}{2\beta m_{i}}}.
\label{wbb14}
\end{eqnarray}
Substituting this expression in Eqs. (\ref{gen7}) and (\ref{gen8}) and
carrying out the integrations, we can show that they return the
results (\ref{diff10}) and (\ref{diff11}) obtained by a different
method. For comparison with the water-bag model, we give below the
asymptotic expressions of the diffusion coefficient and friction force
for a thermal bath. For $v\rightarrow +\infty$, we have
\begin{eqnarray}
D_{\|}=A_{2}\frac{\rho}{\beta}\frac{1}{v^{3}},
\label{wbb15}
\end{eqnarray}
\begin{eqnarray}
D_{\perp}=A_{2}\rho m_{f} \frac{1}{v},
\label{wbb16}
\end{eqnarray}
\begin{eqnarray}
\eta=A_{2}\rho m \frac{1}{v^{2}}.
\label{wbb17}
\end{eqnarray}
For $v\rightarrow 0$, we get
\begin{eqnarray}
D_{\|}=\frac{\pi}{2}A_{2}\left (\frac{\beta}{2\pi}\right )^{1/2}\rho m_{f}^{3/2}\left (1-\frac{3}{8}\beta m_{f}v^{2}+...\right ),\nonumber\\
\label{wbb18}
\end{eqnarray}
\begin{eqnarray}
D_{\perp}=\frac{\pi}{2}A_{2}\left (\frac{\beta}{2\pi}\right )^{1/2}\rho m_{f}^{3/2}\left (1-\frac{1}{8}\beta m_{f}v^{2}+...\right ),\nonumber\\
\label{wbb19}
\end{eqnarray}
\begin{eqnarray}
\eta=\frac{\pi}{2}A_{2}\left (\frac{\beta}{2\pi}\right )^{1/2}\rho m_{f}^{3/2}\beta m v.
\label{wbb20}
\end{eqnarray}
Interestingly, these are the same asymptotic behaviors (same scaling) as in the water-bag model. This is also the case in $d=3$ (see Sec. \ref{sec_d3}).

\subsection{Example: Coulombian plasma}
\label{sec_cp}

In a recent paper, Benedetti {\it et al.} \cite{ben} have
considered a 2D model of Coulomb oscillators interacting via a potential
 of the form
\begin{eqnarray}
\label{cp1}
u_{ij}=-\frac{\xi}{N}\ln |{\bf r}_{i}-{\bf r}_{j}|.
\end{eqnarray} 
They have calculated the friction force ${\bf K}=\langle (\Delta {\bf
v})\rangle/\Delta t$ experienced by a test particle by evaluating the
average variation of the velocity caused by a succession of binary
encounters. The calculations are relatively difficult because the
explicit expression of the differential cross section for a Coulombian
interaction is not known in $d=2$. We shall reconsider this problem
with our approach based on the Landau or Lenard-Balescu equation which
does not require the explicit expression of the differential cross
section, but only the Fourier transform of the potential of
interaction. As noted after Eq. (\ref{rosen3}), for weakly interacting
systems, the form of potential just determines the timescale of
relaxation through the constant $A$.

First, we note that in $d=2$ the potential (\ref{cp1}) is solution of
the differential equation 
\begin{eqnarray}
\label{cp2}
\Delta u=-\frac{2\pi\xi}{N}\delta ({\bf r}). 
\end{eqnarray}
This immediately leads to 
 \begin{eqnarray}
\label{cp3}
(2\pi)^{2}\hat{u}(k)=\frac{2\pi\xi}{Nk^{2}}.
\end{eqnarray}
Using Eq. (\ref{cp3}), we find that the constant (\ref{gen14}) is given by
\begin{eqnarray}
\label{cp4}
A_{2}=\frac{2\pi\xi^{2}}{N^{2}}\Lambda
\end{eqnarray}
where
\begin{eqnarray}
\label{cp5}
\Lambda=\int_{0}^{+\infty}\frac{dk}{k^{2}}.
\end{eqnarray}
We note that for a Coulombian potential in $d=2$, the Coulomb factor
$\Lambda$ diverges {\it linearly} with the distance for
$\lambda=2\pi/k\rightarrow +\infty$. This contrasts from the 3D case
where the divergence is only logarithmic. In paper \cite{hb2}, we have
suggested that this divergence would be cured (as in the 3D case) by
using the complete form of the Lenard-Balescu equation including the
collective effects encapsulated in the dielectric function. These
calculations are completed in Appendix \ref{sec_reg}. However, in
a first approach, we shall neglect collective effects and argue
phenomenologically that $\Lambda$ should scale like the Debye length
in 2D. Note that there can be a numerical factor between $\Lambda$ and
the Debye length so that our approach will only provide an estimate of
the diffusion coefficient.

If we specialize on the case where the distribution of the bath is a step 
function, we can use the results of Sec. \ref{sec_wb} with
\begin{eqnarray}
\label{cp6}
\eta_{0}=\frac{N}{(2\pi)^{2}k_{B}TR^{2}},\qquad v_{0}=2\sqrt{k_{B}T},
\end{eqnarray}
where, following \cite{ben}, we have introduced a typical radius via
$\rho=N/(\pi R^{2})$ and defined the ``temperature'' via $\langle
v^{2}\rangle=2k_{B}T$ (we have taken $m=1$). Recalling that for
identical particles the friction force is given by $K=2\eta$ and using
the asymptotic results (\ref{wbb10}) and (\ref{wbb13}), we get for
$v\rightarrow +\infty$:
\begin{eqnarray}
\label{cp8}
K=4\frac{\xi^{2}}{NR^{2}}\Lambda \frac{1}{v^{2}},
\end{eqnarray}
and for $v\rightarrow 0$:
\begin{eqnarray}
\label{cp9}
K=\frac{1}{2}\frac{\xi^{2}}{NR^{2}}\frac{\Lambda}{(k_{B}T)^{3/2}}v.
\end{eqnarray}
These asymptotic results agree with those obtained by Benedetti {\it
et al.} \cite{ben} with a different method, up to a factor
$2(\pi-2)/\pi\simeq 0.727...$. Owing to the remark following
Eq. (\ref{cp5}), we should not give too much credit on the exact value
of the numerical constant (see Appendix \ref{sec_reg} for a more
precise determination of the value of $\Lambda$). However, it would be
important in future works to determine whether our approach and that
of \cite{ben} are really equivalent or not.

\begin{figure}
\centering
\includegraphics[width=8cm]{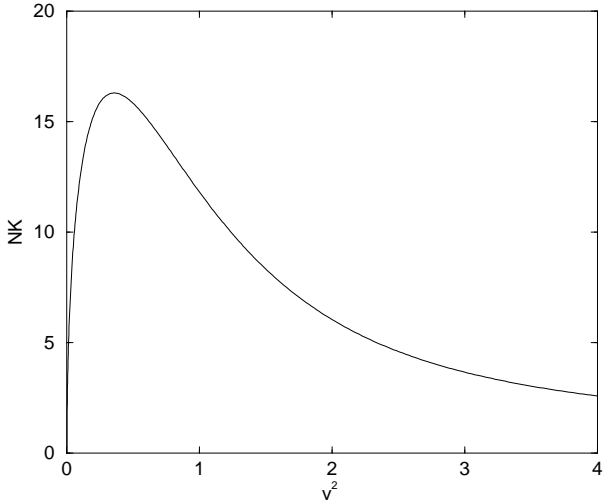}
\caption{Friction force as a function of the velocity for an isothermal distribution of the bath (theoretical curve obtained with our approach). For comparison, we have adopted the same normalization as in \cite{ben}.}
\label{etabenisotherme}
\end{figure}

\begin{figure}
\centering
\includegraphics[width=8cm]{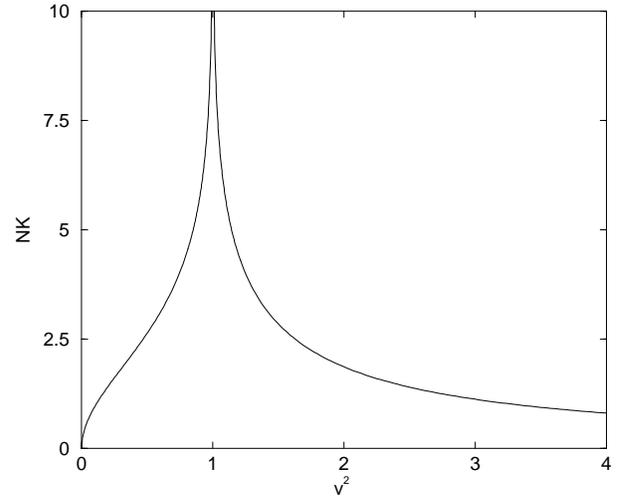}
\caption{Friction force as a function of the velocity when the distribution of the bath is a step function (theoretical curve obtained with our approach). For comparison, we have adopted the same normalization as in \cite{ben}. For $N\rightarrow +\infty$, the friction force diverges logarithmically at the velocity $v_{0}$.}
\label{etabenwater}
\end{figure}

We emphasize that our approach provides the expression of the
diffusion tensor $D^{\mu\nu}$ and friction force $\eta^{\mu}$ for any
value of the velocity; these quantities have not been calculated in
\cite{ben}. When the distribution of the field particles is Maxwellian
(thermal bath), they are given by the analytical formulae
(\ref{diff10}) and (\ref{diff11}) obtained in \cite{hb2}. The friction
force obtained from Eq. (\ref{einstein7}) is plotted as a function of
the velocity in Fig. \ref{etabenisotherme}. We have multiplied our
expression by the factor $2(\pi-2)/\pi$. In that case, our analytical
formula of $K$ matches remarkably well, in the whole range of
velocities, the curve obtained numerically by \cite{ben}. For a
water-bag distribution of the bath, the diffusion coefficients and
friction force can be expressed in terms of integrals of Bessel
functions (\ref{wbb3})-(\ref{wbb7}) or, equivalently, in terms of
Hypergeometric functions using Eqs. (\ref{afin1}), (\ref{afin2}) and
(\ref{afin8}).  The friction force obtained from Eq. (\ref{wbb7}) is
plotted as a function of the velocity in
Fig. \ref{etabenwater}. Again, we have multiplied our expression by
the factor $2(\pi-2)/\pi$ and we get an excellent agreement with the
numerical results of \cite{ben}. We note that in our approach valid
for $N\rightarrow +\infty$, the friction force {\it diverges} for
$v\rightarrow v_{0}$. Using Eq. (\ref{divfric1}), we find that the
divergence is like
\begin{eqnarray}
\label{divfric2}
K\sim 2\frac{\xi^{2}}{NR^{2}}\frac{\Lambda}{k_{B}T}\left (a-b\ln \left |1-\frac{v}{2\sqrt{k_{B}T}}\right |\right),
\end{eqnarray}
where $a=(2-2\gamma-\ln 2-2\psi(3/2))/\pi\simeq 0.025287$ and
$b=1/\pi\simeq 0.3183$. This behavior is consistent with the results
of \cite{ben} who obtain the $K-v$ curve from $N$-body simulations and
find that the maximum velocity increases as $N$ increases. For finite
$N$ systems, there is no singularity. The singularity appears for
$N\rightarrow +\infty$. Formally, the curve of Fig. \ref{etabenwater}
and the divergence of the friction force at $v=v_{0}$ when
$N\rightarrow +\infty$ share some analogies with the divergence of the
specific heats in second order phase transitions (e.g., the curve of
Fig. \ref{etabenwater} looks similar to the $\lambda$-transition in
superfluid Helium).

\subsection{Statistics of fluctuations}
\label{sec_statistics}

In order to better understand the previous results, and in particular
the linear divergence of the diffusion coefficient, we can analyze the
statistics of fluctuations of the force produced by a random
distribution of charges in $d=2$. The force created by the charges at
some point of origin is given by
\begin{eqnarray}
\label{s1}
{\bf F}=\sum_{i=1}^{N}\frac{\xi}{N}\frac{{\bf r}_{i}}{r_{i}^{2}}.
\end{eqnarray}
We shall assume that the charges are randomly distributed with uniform
distribution. In that case, the force ${\bf F}$ fluctuates and we have
to determine its distribution $W({\bf F})$. Formally, this
mathematical problem is the same as the one considered by Chavanis \&
Sire \cite{cs} in their investigation of the statistics of the
fluctuations of velocity ${\bf V}$ created by a random distribution of
point vortices in $d=2$. Therefore, we can apply their results just by
making the substitution ${\bf F}\leftrightarrow {\bf V}$ (we also note
that to leading order $\langle F^{2}\rangle\sim N\times (1/N^{2})\sim 1/N$ so
that the proper scaled variable is $x=F N^{1/2}$). To have exactly the
same notations, we set $\xi/N=\gamma/(2\pi)$.  It is shown in
\cite{cs} that the distribution of ${\bf F}$ is a marginal Gaussian
distribution
\begin{equation}
W({\bf  F})= {4\over  {n\gamma^{2}} \ln N} e^{-{4\pi\over n {\gamma^{2}}\ln N}
{F}^{2}}\quad (F\ll F_{crit}(N)),
\label{s2}
\end{equation} 
\begin{equation}
W({\bf F})= {n\gamma^{2}\over 4 \pi^{2}F^{4}}\quad (F\gg F_{crit}(N)),
\label{s3}
\end{equation}
\begin{equation}
F_{crit}(N)\sim \biggl ({n\gamma^{2}\over 4\pi}\ln N\biggr )^{1/2}\ln^{1/2}(\ln
N),
\label{s4}
\end{equation}
where $n$ is the spatial distribution of the particles assumed to be
homogeneous. This distribution is intermediate between Normal and
L\'evy laws: the core of the distribution is Gaussian but the tail
decreases algebraically as $F^{-4}$. It is dominated by  the
contribution of the nearest neighbor \cite{cs}. Note that the variance of the scaled variable $x=FN^{1/2}$ diverges
logarithmically with $N$ since
\begin{equation}
\langle F^{2}\rangle = {n {\gamma^{2}} \over 4\pi} \ln N. 
\label{s5}
\end{equation}
Analytical results for the spatial correlations of the force are
derived in \cite{csfluid}. Here, we shall concentrate on the
basics in order to make the connection with the kinetic theory
developed previously. If we neglect collective effects, the spatial
correlation function of the force can be written (see Eq. (100) of
Ref. \cite{hb1}):
\begin{equation}
\langle {\bf F}({\bf 0})\cdot {\bf F}({\bf r})\rangle=(2\pi)^{2}n\int k^{2}\hat{u}(k)^{2}e^{-i{\bf k}\cdot {\bf r}}d{\bf k}. 
\label{s6}
\end{equation}
Introducing a polar system of coordinates with the $x$ axis in the direction of ${\bf r}$, using $(2\pi)^{2}\hat{u}(k)=\gamma/k^{2}$ and introducing a large scale cut-off $\Lambda$, we obtain 
\begin{equation}
\langle {\bf F}({\bf 0})\cdot {\bf F}({\bf r})\rangle=\frac{n\gamma^{2}}{2\pi}\int_{1/\Lambda}^{+\infty}\frac{J_{0}(kr)}{k}dk,
\label{s7}
\end{equation}
which behaves like 
\begin{equation}
\langle {\bf F}({\bf 0})\cdot {\bf F}({\bf r})\rangle\simeq \frac{n\gamma^{2}}{2\pi}\ln\left (\frac{\Lambda}{r}\right ),
\label{s8}
\end{equation}
for $r/\Lambda\rightarrow 0$. An alternative derivation of this result
is given in Ref. \cite{csfluid}, Appendix D. In particular, we find
that the correlation function diverges logarithmically for
$r\rightarrow 0$, in agreement with the result (\ref{s5}) [note that
in $d=2$ the length scales as $L\sim (N/n)^{1/2}$ which yields the factor
$\frac{1}{2}\ln N$]. We shall now see how collective effects can
remove this logarithmic divergence.  The expression (\ref{s6})
neglects the correlations between particles. Now, applying Eq. (51) of
Ref. \cite{hb1} to the present context, it is found that the spatial
correlation function is solution of the differential equation
\begin{equation}
\Delta h-k_{D}^{2}h=\beta\gamma\delta({\bf x}),
\label{s9}
\end{equation}
where $k_{D}=(\beta n\gamma)^{1/2}$ is the Debye wavenumber. The spatial correlation is then given by
\begin{equation}
h(x)=-\frac{\beta\gamma}{2\pi}K_{0}(k_{D}x),
\label{s10}
\end{equation}
which is the Debye-H\"uckel result in 2D. The Fourier transform of the
correlation function can be written
\begin{equation}
(2\pi)^{2}n\hat{h}(k)=-\frac{k_{D}^{2}}{k^{2}+k_{D}^{2}},
\label{s11}
\end{equation}
and the correlational energy (see Eq. (57) of paper \cite{hb1}) can be
written $W_{c}=-n\gamma V/(4\pi)\lbrack \gamma_{E}+(1/2)\ln (\beta
n\gamma)-\ln 2\rbrack$ where $\gamma_{E}=0.577...$ is the Euler
constant. If we account for collective effects in the computation of
the force auto-correlation function, we obtain (see Eq. (96) of
Ref. \cite{hb1}):
\begin{equation}
\langle {\bf F}({\bf 0})\cdot {\bf F}({\bf r})\rangle=(2\pi)^{2}n\int \frac{k^{2}\hat{u}(k)^{2}}{1+(2\pi)^{2}\beta n \hat{u}(k)}e^{-i{\bf k}\cdot {\bf r}}d{\bf k},
\label{s12}
\end{equation}
where the new terms arise because of the correlations encapsulated in
the function $\hat{h}(k)$. Thus, combining the previous results, we get
\begin{equation}
\langle {\bf F}({\bf 0})\cdot {\bf F}({\bf r})\rangle=\frac{n\gamma^{2}}{2\pi}\int_{0}^{+\infty}\frac{J_{0}(kr)}{k}\frac{k^{2}}{k^{2}+k_{D}^{2}}dk,
\label{s13}
\end{equation}
where there is no need to introduce an {\it ad hoc} large-scale cut-off anymore. We now obtain the result
\begin{equation}
\langle {\bf F}({\bf 0})\cdot {\bf F}({\bf r})\rangle=\frac{n\gamma^{2}}{2\pi}K_{0}(k_{D}r).
\label{s14}
\end{equation} 
For $k_{D}r\rightarrow 0$, we recover Eq. (\ref{s8}) with
$\Lambda=k_{D}^{-1}$. Therefore, the divergence in Eq. (\ref{s7}) can
be regularized by properly accounting for correlations among the
particles. For comparison, the spatial correlation of the force in 3D
within the mean-field Debye-H\"uckel theory is
\begin{eqnarray}
\langle {\bf F}({\bf 0})\cdot {\bf F}({\bf r})\rangle=n\gamma^{2}\int_{0}^{+\infty}\frac{\sin(kr)}{kr}\frac{k^{2}}{k^{2}+k_{D}^{2}}dk\nonumber\\
=\frac{\pi}{2}n\gamma^{2}\frac{1}{r}e^{-k_{D}r},\qquad\qquad 
\label{s15}
\end{eqnarray}
where $k_{D}=(4\pi\beta n e^{2})^{1/2}$ is the Debye shielding length
in $d=3$.  Note that the integral remains well-behaved if we neglect
collective effects ($k_{D}=0$) contrary to the case $d=2$. On the
other hand, the fact that the correlation function diverges as $1/r$
for $r\rightarrow 0$ is due to the fact that the distribution $W({\bf
F})$ is a particular L\'evy law, called the Holtzmark distribution,
whose variance is infinite \cite{channeu}.

Let us now consider the temporal correlations of the force. If we
neglect collective effects, the correlation function is given by (see
Eq. (93) of Ref. \cite{hb2}):
\begin{equation}
\langle {\bf F}(0)\cdot {\bf F}(t)\rangle=(2\pi)^{2}n\int k^{2}\hat{u}(k)^{2}e^{-i{\bf k}\cdot {\bf v}t}e^{-k^{2}t^{2}/(2\beta)}d{\bf k}. 
\label{s16}
\end{equation}
Using the same procedure as before, we get
\begin{equation}
\langle {\bf F}(0)\cdot {\bf F}(t)\rangle=\frac{n\gamma^{2}}{2\pi}\int_{vt/\Lambda}^{+\infty}\frac{J_{0}(x)}{x}e^{-x^{2}/(2\beta v^{2})}dx,  
\label{s17}
\end{equation}
which behaves like 
\begin{equation}
\langle {\bf F}(0)\cdot {\bf F}(t)\rangle=\frac{n\gamma^{2}}{2\pi}\ln\left (\frac{\Lambda}{vt}\right ),
\label{s18}
\end{equation}
for $vt/\Lambda\rightarrow 0$. The previous expression indicates that
the correlation of the force is almost independent of time. This
implies that the diffusion coefficient, calculated with the Kubo
formula
\begin{equation}
D=\int_{0}^{+\infty} \langle {\bf F}(0)\cdot {\bf F}(t)\rangle dt\sim \int_{0}^{+\infty} dt, 
\label{s19}
\end{equation}
diverges {\it linearly} with time. This temporal divergence is the analogue of
the spatial divergence $\Lambda\sim\int_{0}^{+\infty}d\lambda$
obtained in the kinetic approach of Sec. \ref{sec_cp}. The same effect occurs 
for gravitational systems  in $d=3$ where the {\it logarithmic} divergence of the diffusion coefficient can be viewed either as a spatial $\sim \int dk/k$ or temporal $\int dt/t$ divergence (see Sec. 2.9 of \cite{hb2}).
Now, if we take into account collective effects, the temporal correlation function is given by (see Eq. (98) of Ref. \cite{hb2}):
\begin{equation}
\langle {\bf F}(0)\cdot {\bf F}(t)\rangle=(2\pi)^{2}\int\frac{ k^{2}\hat{u}(k)^{2}}{|\epsilon({\bf k},{\bf k}\cdot {\bf v}_{1})|^{2}}e^{i{\bf k}\cdot {\bf u}t}f({\bf v}_{1})d{\bf v}_{1}d{\bf k}. 
\label{s20}
\end{equation}
By investigating the poles of the integrand (for a Maxwellian
distribution), it is found that the contribution of each mode $k$
should decay with time exponentially rapidly (modulated by an
oscillatory factor) with an exponent $\gamma_{k}$ corresponding to
the Landau damping rate. This is the imaginary part of the pulsation
$\omega$ which cancels out the dielectric function $\epsilon({\bf
k},\omega)$. Therefore, collective effects modify the expression of
the temporal correlation function and, consequently, of the diffusion
coefficient. Indeed, according to the Kubo formula, the diffusion
coefficient entering in the Lenard-Balescu equation
\begin{equation}
D^{\mu\nu}=\pi (2\pi)^{2}\int d{\bf v}_{1}d{\bf k} k^{\mu}k^{\nu}\frac{\hat{u}(k)^{2}}{|\epsilon({\bf k},{\bf k}\cdot {\bf v}_{1})|^{2}}\delta({\bf k}\cdot {\bf u})f({\bf v}_{1}), 
\label{s21}
\end{equation}
is the time integral of the correlation function $\langle
F^{\mu}(0)F^{\nu}(t)\rangle$. When the dielectric function is taken
into account, it is seen that the integrals on $k$ in Eq. (\ref{s21})
are now convergent (see discussion in Sec. 2.8.1 of Ref. \cite{hb2}).
By these means, we can provide a justification of the upper
cut-off which appears in Eq. (\ref{cp5}). The calculations are detailed in Appendix \ref{sec_reg}.

\section{The case $d=1$}
\label{sec_d1}

For $d=1$, the Fokker-Planck equation (\ref{fp1}) reduces to
\begin{eqnarray}
\label{un1}
\frac{\partial P}{\partial t}=(2\pi)^{2}\frac{\partial}{\partial v}\int_{-\infty}^{+\infty} d{v}'\int_{0}^{+\infty} d{k} \ k^{2}\frac{\hat{u}(k)^{2}}{|\epsilon({k},{k}{v})|^{2}}\delta\lbrack k(v-v')\rbrack\nonumber\\
\times \sum_{j} \left (m_{j}f'_{j}\frac{\partial P}{\partial v}-m P \frac{d f'_{j}}{d v^{'}}\right ).\qquad\qquad
\end{eqnarray} 
Using $\delta\lbrack k(v-v')\rbrack=(1/|k|)\delta(v-v')$, the integral over $v'$ is straightforward and yields
\begin{eqnarray}
\label{un2}
\frac{\partial P}{\partial t}=\frac{\partial}{\partial v}A(v)\sum_{j} \left (m_{j}f_{j}\frac{\partial P}{\partial v}-m P \frac{d f_{j}}{d v}\right ),
\end{eqnarray} 
where 
\begin{eqnarray}
\label{un3}
A(v)=(2\pi)^{2}\int_{0}^{+\infty} d{k} \frac{k\hat{u}(k)^{2}}{|\epsilon({k},{k}{v})|^{2}}.
\end{eqnarray} 
This can be rewitten in the form
\begin{eqnarray}
\label{un4}
\frac{\partial P}{\partial t}=\frac{\partial}{\partial v}\left (D\frac{\partial P}{\partial v}+P\eta\right ),
\end{eqnarray} 
with
\begin{eqnarray}
D(v)=A(v)\sum_{j}m_{j}f_{j}(v),
\label{un5}
\end{eqnarray}
and 
\begin{eqnarray}
\eta(v)=-m A(v)\sum_{j}\frac{df_{j}}{dv}.
\label{un6}
\end{eqnarray}
If the field particles all have the same mass $m_{f}$, these
expressions simplify in
\begin{eqnarray}
D(v)=A(v) m_{f}f(v), \qquad \eta(v)=-A(v) m f'(v).
\label{un7}
\end{eqnarray}
The effective potential (\ref{iso7}) is
\begin{eqnarray}
U(v)=-\frac{m}{m_{f}}\ln f(v),
\label{un8}
\end{eqnarray}
and the Fokker-Planck equation can be rewritten
\begin{eqnarray}
\frac{\partial P}{\partial t}=\frac{\partial}{\partial v}\left \lbrack D(v)\left (\frac{\partial P}{\partial v}-P\frac{m}{m_{f}}\frac{d \ln f}{d v}\right )\right\rbrack,
\label{un9}
\end{eqnarray}
with
\begin{eqnarray}
D(v)=(2\pi)^{2}m_{f}f(v)\int_{0}^{+\infty} \frac{k\hat{u}(k)^{2}}{|\epsilon({k},{k}{v})|^{2}} d{k}.
\label{un10}
\end{eqnarray}
We note that the stationary solution of the Fokker-Planck equation (\ref{un9}) is
\begin{eqnarray}
P^{eq}(v)=C\ f^{\frac{m}{m_{f}}}(v),
\label{un11}
\end{eqnarray}
and that $P^{eq}(v)=f(v)$ when the mass of the test particle is equal to the mass of the field particles $m=m_{f}$. More explicit expressions of the diffusion coefficient valid for a thermal bath are given in \cite{hb2}.

\section{Conclusion}
\label{sec_concl}

In this paper, we have discussed the kinetic theory of systems with
long-range interactions in a relatively unified framework starting
from the Landau and Lenard-Balescu equations. Using a test particle
approach, we have given explicit expressions for the diffusion
coefficient and friction force entering in the Fokker-Planck equation
for different potentials of interaction in different dimensions of
space and for different distributions of the bath. We have also
considered the possibility of a distribution of mass among the particles and
shown how the results (in particular the Einstein relation) are
modified in that case.

For Coulombian and Newtonian potentials in $d=3$, we have enlightened
the connection between results of plasma physics and results of
stellar dynamics which have been derived almost independently and in a
relatively different form. We have shown that the kinetic equation
derived in stellar dynamics by Chandrasekhar \cite{chandra} and
Rosenbluth {\it et al.} \cite{rosen} from the Fokker-Planck equation
can also be obtained from the Landau \cite{landau} equation of plasma
physics. We have then considered the extension of these kinetic
theories in $d=2$.  For a Maxwellian distribution of the bath, the
diffusion and friction coefficients can be calculated explicitly in
terms of Bessel functions as shown in \cite{hb2}. In the present
paper, we have generalized our approach to an arbitrary isotropic
distribution of the bath and expressed the results in terms of
integrals of Bessel functions. In the case where the distribution
function of the field particles is a step function, we have shown that
the asymptotic expressions of our obtained diffusion and friction
coefficients reproduce those found by Benedetti {\it et al.} 
\cite{ben} (up to a factor $2(\pi-2)/\pi$)
with a different method. More generally, we get a good agreement with
their numerical results for all velocities. We have shown analytically
that, for $N\rightarrow +\infty$, the friction diverges
logarithmically at the critical velocity $v_{0}$ (while for finite $N$
there is no divergence). We have also shown that the linear divergence
of the diffusion coefficient resulting from the Landau approximation
could be removed by considering collective effects as in the
Lenard-Balescu treatment of a 3D plasma. Finally, we have shown how the
results of kinetic theory simplify in $d=1$.

For future perspectives, it could be mentioned that the results
presented in this paper can be formally extended to a generalized class of
kinetic equations, associated with a generalized thermodynamical
framework, introduced by Kaniadakis \cite{kaniadakis} and Chavanis
\cite{genpre,genlandau}. 
These generalized equations could be justified in the case of complex
systems when the transition probabilities from one state to the other
are different from the form that is usually considered due to
microscopic constraints (``hidden constraints'') that affect the
dynamics. However, the domains of application of these generalized
kinetic theories remains to be better specified.

\appendix

\section{Alternative derivation of the diffusion coefficients and friction force in $d=3$}
\label{sec_alternative}

In Sec. \ref{sec_d3}, we have derived the expressions of the diffusion
coefficients and friction force for an isotropic velocity distribution
of the field particles in $d=3$ by using the Rosenbluth potentials. In
this Appendix, we show that we can obtain the same results by using
the method developed in Sec. \ref{sec_d2} which extends to $d=2$.

Starting from the general expression (\ref{gen4}) of the diffusion
coefficient and introducing a spherical system of coordinates with the
$z$-axis in the direction of ${\bf v}$, we find after straightforward
algebra, that the diffusion tensor can be written as in Eq. (\ref{iso1}) with
\begin{eqnarray}
D_{\|}=(2\pi)^{7}\sum_{j}m_{j}\frac{1}{v}\int_{0}^{+\infty}k^{3}\hat{u}(k)^{2}dk\nonumber\\
\times \int_{0}^{+\infty}dx \left\lbrack 2\left ( 1-\frac{2}{x^{2}}\right )\frac{\sin x}{x}+4\frac{\cos x}{x^{2}}\right\rbrack\Phi_{j}\left (\frac{x}{v}\right),\quad
\label{alt1}
\end{eqnarray}
\begin{eqnarray}
D_{\perp}=(2\pi)^{7}\sum_{j}m_{j}\frac{1}{v}\int_{0}^{+\infty}k^{3}\hat{u}(k)^{2}dk\nonumber\\
\times \int_{0}^{+\infty}dx \left\lbrack 4\frac{\sin x}{x^{3}}-4\frac{\cos x}{x^{2}}\right\rbrack\Phi_{j}\left (\frac{x}{v}\right).
\label{alt2}
\end{eqnarray}
The Fourier transform of an isotropic velocity distribution of the
field particles in $d=3$ is
\begin{eqnarray}
\Phi(k)=\frac{1}{2\pi^{2}k}\int_{0}^{+\infty}f(v_{1})\sin (k v_{1})v_{1} dv_{1}.
\label{alt3}
\end{eqnarray}
Substituting this expression in Eqs. (\ref{alt1}) and (\ref{alt2}) and
introducing the functions
\begin{eqnarray}
I_{\|}(\lambda)=\int_{0}^{+\infty}dx  \left\lbrack 2\left ( 1-\frac{2}{x^{2}}\right )\frac{\sin x}{x}+4\frac{\cos x}{x^{2}}\right\rbrack\frac{1}{x}\sin(\lambda x),\nonumber\\
\label{alt4}
\end{eqnarray}
\begin{eqnarray}
I_{\perp}(\lambda)=\int_{0}^{+\infty}dx  \left\lbrack 4\frac{\sin x}{x^{3}}-4\frac{\cos x}{x^{2}}\right\rbrack\frac{1}{x}\sin(\lambda x),\nonumber\\
\label{alt5}
\end{eqnarray}
we obtain
\begin{eqnarray}
D_{\|}=8A\sum_{j}m_{j}\int_{0}^{+\infty} f_{j}(v_{1})I_{\|}\left (\frac{v_{1}}{v}\right )v_{1}dv_{1}. 
\label{alt6}
\end{eqnarray}
\begin{eqnarray}
D_{\perp}=8A\sum_{j}m_{j}\int_{0}^{+\infty} f_{j}(v_{1})I_{\perp}\left (\frac{v_{1}}{v}\right )v_{1}dv_{1}. 
\label{alt7}
\end{eqnarray}
The integrals (\ref{alt4}) and (\ref{alt5}) can be calculated explicitly. Using the results:  $I_{\|}(\lambda)=\pi/3$, $I_{\perp}(\lambda)=2\pi/3$ for $\lambda>1$ and $I_{\|}(\lambda)=(\pi/3)\lambda^3$, $I_{\perp}(\lambda)=\pi \lambda-(\pi/3)\lambda^{3}$ for $\lambda<1$, we recover the expressions (\ref{di5}) and (\ref{di6}). The friction $\eta$ can be obtained from Eq. (\ref{iso3}) and we recover Eq. (\ref{di8}).

\section{Asymptotic behaviors of the functions (\ref{wbb5}) and (\ref{wbb6}) }
\label{sec_asymptotic}

\subsection{Asymptotic behaviors for $\lambda\rightarrow 0$}
\label{sec_a1}

In this Appendix, we determine the asymptotic behaviors of the
functions (\ref{wbb5}) and (\ref{wbb6}) for $\lambda\rightarrow
0$. Using identities (\ref{gen5}) and (\ref{gen6}), Eq. (\ref{wbb5})
can be rewritten
\begin{eqnarray}
R_{\|}(\lambda)=\frac{1}{\pi}\int_{0}^{\pi}d\theta \cos^{2}\theta \int_{0}^{+\infty} e^{ix\cos\theta} \frac{J_{1}(\lambda x)}{x}dx.
\label{aun1}
\end{eqnarray}
Setting $t=\cos\theta$, we obtain
\begin{eqnarray}
R_{\|}(\lambda)=\frac{1}{\pi} \int_{-1}^{+1} \frac{t^{2}dt}{\sqrt{1-t^{2}}}\int_{0}^{+\infty}  e^{ixt}  \frac{J_{1}(\lambda x)}{x}dx.
\label{aun2}
\end{eqnarray}
In this expression, $t$ and $x$ are real and the domains of
integration $\tau_{0}: -1\le t \le 1$ and $\zeta_{0}: 0\le x<+\infty$
are taken along the real axis. Under these circumstances, we cannot
simply expand the last function in Taylor series for
$\lambda\rightarrow 0$ because the resulting integrals would not be
convergent. However, regarding $x$ and $t$ as complex variables, it is
possible to choose paths of integration along which this expansion
will converge. We shall first carry out the integration on $x$, for a
fixed $t$. It will therefore be possible to choose the (complex)
integration paths for $x$ dependent on $t$. The integration paths are
modified as follows: $\tau_{0}$ is replaced by $\tau$, the semi-circle
with radius unity lying in the domain ${\cal I}_{m}(t)\ge
0$. Therefore, $\arg(t)$ varies from $\pi$ to $0$ when $t$ moves from
$-1$ to $+1$. On the other hand, $\zeta_{0}$ is replaced by
$\zeta_{\psi_{t}}$, the line starting from the origin and forming an
angle
\begin{equation}
\psi_{t}={\pi\over 2}-\arg (t)
\label{aun3}
\end{equation}      
with the real axis. With these new contours, 
\begin{eqnarray}
R_{\|}(\lambda)=\frac{1}{\pi}{\rm Re} \int_{\tau} \frac{t^{2}dt}{\sqrt{1-t^{2}}}\int_{\zeta_{\psi_{t}}} e^{ixt}  \frac{J_{1}(\lambda x)}{x}dx.
\label{aun4}
\end{eqnarray}
When $t$ moves from $-1$ to $+1$, $\psi_{t}$ varies from $-{\pi\over
2}$ to $+{\pi\over 2}$. The argument of $i x t$ is equal to
$\pi$. Therefore, the real part of $i x t$ is always negative and the
function $e^{i x t}$ decays exponentially to zero as
$|x|\rightarrow\infty$. Therefore, with the new paths of integration
$\tau$ and $\zeta_{\psi_{t}}$, it is possible to expand the
integrand of Eq. (\ref{aun4}) in power series of $\lambda$, for
$\lambda\rightarrow 0$, and integrate term by term:
\begin{eqnarray}
R_{\|}(\lambda)=\frac{1}{\pi}{\rm Re} \int_{\tau} \frac{t^{2}dt}{\sqrt{1-t^{2}}}\int_{\zeta_{\psi_{t}}} e^{ixt}  \left (\frac{\lambda}{2}-\frac{\lambda^{3}x^{2}}{16}+...\right )dx.\nonumber\\
\label{aun5}
\end{eqnarray}
Setting $ i x t =-y$ ($y \ {\rm
real}\ge 0 $), we get
\begin{eqnarray}
R_{\|}(\lambda)=-\frac{1}{\pi}{\rm Re} \int_{-1}^{+1}\frac{t^{2}dt}{\sqrt{1-t^{2}}}\nonumber\\
\times\int_{0}^{+\infty} e^{-y}  \left (\frac{\lambda}{2}+\frac{\lambda^{3}y^{2}}{16t^{2}}+...\right )\frac{dy}{it},
\label{aun6}
\end{eqnarray}
where we recall that $t$ is a complex variable and the integration has 
to be performed over the semi-circle of radius unity lying on the domain
${\cal I}_{m}(t)\ge 0$. Writing $t=e^{i\theta}$, we find that
\begin{eqnarray}
{\rm Re}\ i \int_{-1}^{+1}\frac{t dt}{\sqrt{1-t^{2}}}=0,
\label{aun7}
\end{eqnarray}
and 
\begin{eqnarray}
{\rm Re}\ i \int_{-1}^{+1}\frac{dt}{t\sqrt{1-t^{2}}}=\pi.
\label{aun8}
\end{eqnarray}
Therefore,
\begin{eqnarray}
R_{\|}(\lambda)\sim \frac{\lambda^{3}}{16}\Gamma(3)=\frac{\lambda^{3}}{8}. 
\label{aun9}
\end{eqnarray}
Using a similar procedure, we find that
\begin{eqnarray}
R_{\perp}(\lambda)\sim \frac{\lambda}{2}. 
\label{aun10}
\end{eqnarray}
Note that this second result can also be obtained directly from
Eq. (\ref{wbb6}) by using $J_{1}(\lambda x)/x\sim \lambda/2$ and
\begin{eqnarray}
\int_{0}^{+\infty}\frac{J_{1}(x)}{x}dx=1.
\label{aun11}
\end{eqnarray}

\subsection{Asymptotic behaviors for $\lambda\rightarrow +\infty$}
\label{sec_a2}

We now determine the asymptotic behaviors of the functions (\ref{wbb5})
and (\ref{wbb6}) for $\lambda\rightarrow +\infty$. Setting $z=\lambda
x$, we can rewrite Eq. (\ref{wbb5}) as
\begin{eqnarray}
R_{\|}(\lambda)=\int_{0}^{+\infty}dz \frac{J_{1}(z)}{z}\left\lbrack J_{0}\left (\frac{z}{\lambda}\right )-\frac{J_{1}\left (\frac{z}{\lambda}\right )}{\frac{z}{\lambda}}\right\rbrack.
\label{ad1}
\end{eqnarray}
Using identity (\ref{gen6}), we obtain
\begin{eqnarray}
R_{\|}(\lambda)=\frac{1}{\pi}\int_{0}^{\pi}d\theta \sin^{2}\theta \int_{0}^{+\infty}dz e^{iz\cos\theta}\nonumber\\
\times \left\lbrack J_{0}\left (\frac{z}{\lambda}\right )-\frac{J_{1}\left (\frac{z}{\lambda}\right )}{\frac{z}{\lambda}}\right\rbrack.\nonumber\\
\label{ad2}
\end{eqnarray}
Setting $t=\cos\theta$, using the contours introduced in  the previous section and expanding the last term in Taylor series for $1/\lambda\rightarrow 0$, we
find that
\begin{eqnarray}
R_{\|}(\lambda)=\frac{1}{\pi}{\rm Re} \int_{\tau}dt \sqrt{1-t^{2}} \int_{\zeta_{\psi_{t}}} e^{izt}\left\lbrack \frac{1}{2}-\frac{3}{16}\frac{z^{2}}
{\lambda^{2}}+...\right\rbrack dz.  \nonumber\\
\label{ad3}
\end{eqnarray}
Setting $izt=-y$ ($y \ {\rm
real}\ge 0 $)  we get
\begin{eqnarray}
R_{\|}(\lambda)=-\frac{1}{\pi}{\rm Re} \int_{-1}^{+1}dt  \sqrt{1-t^{2}}  \nonumber\\
\times \int_{0}^{+\infty} e^{-y}\left\lbrack \frac{1}{2}+\frac{3}{16}\frac{y^{2}}{\lambda^{2}t^{2}}+...\right\rbrack \frac{dy}{it}, 
\label{ad4}
\end{eqnarray}
where the integration on $t$ has to be carried out on the upper semi-circle in the complex plane. Using the identities
\begin{eqnarray}
{\rm Re}\ i \int_{-1}^{+1}\sqrt{1-t^{2}}\frac{dt}{t}=\pi,
\label{ad5}
\end{eqnarray}
\begin{eqnarray}
{\rm Re}\ i \int_{-1}^{+1}\sqrt{1-t^{2}}\frac{dt}{t^{3}}=-\frac{\pi}{2},
\label{ad6}
\end{eqnarray}
we find that
\begin{eqnarray}
R_{\|}(\lambda)=\frac{1}{2}-\frac{3}{16\lambda^{2}}+...
\label{ad7}
\end{eqnarray}
Using a similar procedure, we obtain
\begin{eqnarray}
R_{\perp}(\lambda)=\frac{1}{2}-\frac{1}{16\lambda^{2}}+...
\label{ad8}
\end{eqnarray}

\subsection{Relation to hypergeometric functions and behavior for $\lambda\sim 1$}
\label{sec_a3}

In fact, the functions (\ref{wbb5}) and (\ref{wbb6}) can be expressed
in terms of Hypergeometric functions. We can then easily obtain their
asymptotic behaviors from  standard formulae \cite{abramowitz}. We have
\begin{eqnarray}
R_{\|}(\lambda)=\frac{\lambda}{2}\left\lbrack F\left (\frac{1}{2},\frac{1}{2},2,\lambda^{2}\right )-F\left (-\frac{1}{2},\frac{1}{2},2,\lambda^{2}\right )\right\rbrack,\nonumber\\
\label{afin1}
\end{eqnarray}
\begin{eqnarray}
R_{\perp}(\lambda)=\frac{\lambda}{2} F\left (-\frac{1}{2},\frac{1}{2},2,\lambda^{2}\right ).
\label{afin2}
\end{eqnarray}
For $\lambda\rightarrow 0$, 
\begin{eqnarray}
R_{\|}(\lambda)=\frac{\lambda^{3}}{8}+\frac{\lambda^{5}}{32}+\frac{15\lambda^{7}}{1024}+...
\label{afin3}
\end{eqnarray}
\begin{eqnarray}
R_{\perp}(\lambda)=\frac{\lambda}{2}-\frac{\lambda^{3}}{16}-\frac{\lambda^{5}}{128}-\frac{5\lambda^{7}}{2048}+...
\label{afin4}
\end{eqnarray}
and for $\lambda\rightarrow +\infty$,
\begin{eqnarray}
R_{\|}(\lambda)=\frac{1}{2}-\frac{3}{16 \lambda^2}-\frac{5}{128\lambda^{4}}+...
\label{afin5}
\end{eqnarray}
\begin{eqnarray}
R_{\perp}(\lambda)=\frac{1}{2}-\frac{1}{16 \lambda^2}-\frac{1}{128\lambda^{4}}+...
\label{afin6}
\end{eqnarray}
We also note the particular values
\begin{eqnarray}
R_{\|}(1)=\frac{2}{3\pi},\qquad R_{\perp}(1)=\frac{4}{3\pi}.
\label{afin7}
\end{eqnarray}
From Eq. (\ref{afin1}), we get
\begin{eqnarray}
R_{\|}'(\lambda)=\frac{1}{2}\left\lbrack F\left (\frac{1}{2},\frac{1}{2},2,\lambda^{2}\right )-F\left (-\frac{1}{2},\frac{1}{2},2,\lambda^{2}\right )\right\rbrack\nonumber\\
+\frac{\lambda^{2}}{8}\left\lbrack F\left (\frac{1}{2},\frac{3}{2},3,\lambda^{2}\right )+F\left (\frac{3}{2},\frac{3}{2},3,\lambda^{2}\right )\right\rbrack.\quad
\label{afin8}
\end{eqnarray}
This function diverges for $\lambda\rightarrow 1$ like
\begin{eqnarray}
R_{\|}'(\lambda)=\frac{4}{3\pi}-\frac{1}{\pi}(2\gamma+\ln 2+\ln |1-\lambda|+2\psi(3/2)),\nonumber\\
\label{afin9}
\end{eqnarray}
where $\gamma=0.577216...$ is the Euler constant and $\psi(3/2)=0.03649...$ is the Digamma function \cite{abramowitz}.

\section{Regularization of the linear divergence}
\label{sec_reg}

In this Appendix, we show how the linear divergence of the diffusion
coefficient for a Coulombian plasma in $d=2$ can be regularized by
taking into account collective effects. When collective effects are
taken into account using Eq. (\ref{fp1}) instead of Eq. (\ref{fp2}),
the diffusion coefficient is given by
\begin{equation}
D^{\mu\nu}=\pi (2\pi)^{d}m\int d{\bf v}_{1}d{\bf k} k^{\mu}k^{\nu}\frac{\hat{u}(k)^{2}}{|\epsilon({\bf k},{\bf k}\cdot {\bf v})|^{2}}\delta({\bf k}\cdot {\bf u})f({\bf v}_{1}). 
\label{reg1}
\end{equation}
We concentrate here on a thermal bath with Maxwellian
distribution. Using the same method as in Sec. \ref{sec_diffusion} but
keeping the dielectric function, we obtain 
\begin{eqnarray}
D^{\mu\nu}=\pi (2\pi)^{d}m\rho \left (\frac{\beta
m}{2\pi}\right)^{1/2}\qquad\qquad\qquad\nonumber\\
\times\int d{\bf k}
\frac{k^{\mu}k^{\nu}}{k}\frac{\hat{u}(k)^{2}}{|\epsilon({\bf k},{\bf
k}\cdot {\bf v})|^{2}}e^{-\beta m\frac{({\bf k}\cdot {\bf v})^{2}}{2
k^{2}}}.
\label{reg2}
\end{eqnarray}
For a thermal bath one has (see, e.g., \cite{hb2}):
 \begin{equation}
|\epsilon({\bf k},{\bf k}\cdot {\bf v})|^{2}=(1-\eta(k)B(\hat{\bf k}\cdot {\bf x}))^{2}+C(\hat{\bf k}\cdot {\bf x})^{2},  
\label{reg3}
\end{equation}
where we have defined $\hat{\bf k}={\bf k}/k$, ${\bf x}=(\beta m/2)^{1/2}{\bf v}$, $\eta(k)=-(2\pi)^{d}\hat{u}(k)\beta m\rho$, $B(z)=1-2 z e^{-z^{2}}\int_{0}^{z}e^{t^{2}}dt$ and $C(z)=\sqrt{\pi}|z|e^{-z^{2}}$. The diffusion tensor can be rewritten
 \begin{equation}
D^{\mu\nu}=\frac{\pi}{(2\pi)^{d}\beta^{2}\rho m} \left (\frac{\beta m}{2\pi}\right)^{1/2}\int d \hat{\bf k} {\hat k}^{\mu}{\hat k}^{\nu} I(\hat{\bf k}\cdot {\bf x}) e^{-(\hat{\bf k}\cdot {\bf x})^{2}}, 
\label{reg4}
\end{equation}
where
 \begin{equation}
I(z)=\int_{0}^{+\infty} \frac{k^{d}\eta(k)^{2} dk}{(1-\eta(k)B(z))^{2}+C(z)^{2}}.
\label{reg5}
\end{equation}
For a Coulombian potential, one has $\eta(k)=-k_{D}^{2}/k^{2}$ and we obtain
 \begin{equation}
I(z)=\int_{0}^{+\infty} \frac{k^{d}dk}{(B(z)+k^{2}/k_{D}^{2})^{2}+C(z)^{2}}.
\label{reg6}
\end{equation}
This can be rewritten
$I(z)=k_{D}^{d+1}C(z)^{\frac{d-3}{2}}\Phi_{d}(B(z)/C(z))$ where
$\Phi_{d}(z)=\int_{0}^{+\infty} t^{d} dt/((z+t^{2})^{2}+1)$. If we
neglect collective effects we have instead 
\begin{equation}
I^{Landau}=k_{D}^{4}\int_{0}^{+\infty}\frac{dk}{k^{4-d}},
\label{reg7}
\end{equation}
which presents divergences for $k\rightarrow 0$ if $d\le 3$. The
regularization of the divergence in $d=3$ has been treated by Balescu
\cite{balescu} and the regularization in $d=1$ has been treated in
\cite{hb2}. Let us focus here on the case $d=2$. We note that, contrary to
Eq. (\ref{reg7}), the integral (\ref{reg6}) is well-behaved for
$k\rightarrow 0$ and $k\rightarrow +\infty$. Therefore, there is no
linear divergence of the diffusion coefficient and friction force when
collective effects are taken into account.

\begin{figure}
\centering
\includegraphics[width=8cm]{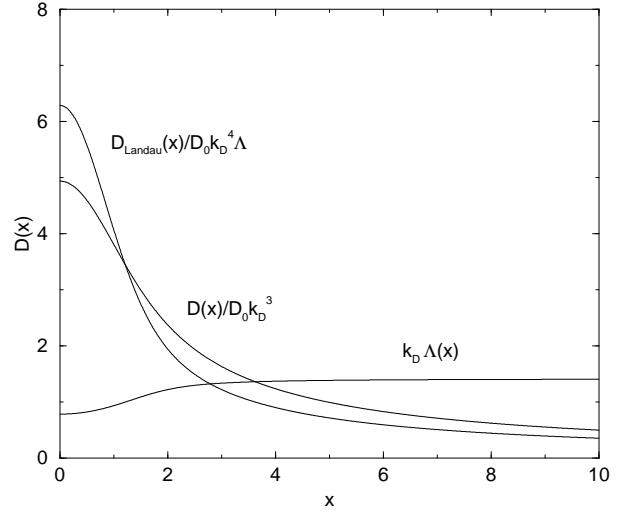}
\caption{Dependence of the trace of the diffusion tensor with the velocity, using (i) the Landau approximation or (ii) the Lenard-Balescu treatment of collective effects. }
\label{interpol}
\end{figure}

To obtain the expression of the diffusion tensor, one has to
substitute Eq. (\ref{reg6}) in Eq. (\ref{reg4}), using
$z=x\cos\theta$, and carry out the integrations. We shall simplify the
calculations a little bit by introducing an approximate analytical
expression of $I(z)$ (we have checked numerically that the exact
treatment yields close results). Let us first consider asymptotic
behaviors of the previously defined functions. We have
$\Phi_{2}(z)=(\pi/(2\sqrt{2}))(1-{z}/{2}+{z^{2}}/{8}+...)$ for
$z\rightarrow 0$, $\Phi_{2}(z)\sim {\pi}/({4\sqrt{z}})$ for
$z\rightarrow +\infty$ and $\Phi_{2}(z)\sim ({\pi}/{2})\sqrt{-z}$ for
$z\rightarrow -\infty$. On the other hand, $B(z)/C(z)\sim
1/(\sqrt{\pi}|z|)$ for $z\rightarrow 0$ and $B(z)/C(z)\sim
-1/(2\sqrt{\pi}|z|^3)e^{z^{2}}$ for $|z|\rightarrow +\infty$. Thus, we
find that $I(z)/k_{D}^{3}\rightarrow \pi/4$ for $z\rightarrow 0$ and
$I(z)/k_{D}^{3}\rightarrow \sqrt{\pi}/(2\sqrt{2}z^{2})e^{z^{2}}$ for
$z\rightarrow +\infty$. Let us consider a simple interpolation formula
of the form
\begin{equation}
\frac{I(z)}{k_{D}^{3}}=\frac{\pi}{4}+\frac{1}{2}\left (\frac{\pi}{2}\right
)^{1/2}\left (\frac{e^{z^{2}}}{z^{2}}-\frac{1}{z^{2}}-1\right ).
\label{reg8}
\end{equation} 
With this expression, it turns out that the trace of the diffusion tensor
\begin{equation}
D^{\mu\mu}=D_{0}\int_{0}^{2\pi}I(x\cos\theta)e^{-x^{2}\cos^{2}\theta}d\theta,
\label{reg9}
\end{equation} 
can be calculated analytically ($D_{0}$ is the value of the constant in front of the integral in Eq. (\ref{reg4})). When we use Eq. (\ref{reg8}), we obtain
\begin{eqnarray}
D^{\mu\mu}=D_{0}k_{D}^{3}\frac{\pi^{3/2}}{2}e^{-\frac{x^{2}}{2}}\left\lbrack \sqrt{\pi}I_{0}\left (\frac{x^{2}}{2}\right )+\sqrt{2}I_{1}\left (\frac{x^{2}}{2}\right )\right\rbrack.\nonumber\\
\label{reg10}
\end{eqnarray}
Alternatively, when we use Eq. (\ref{reg7}), we get
\begin{equation}
D_{Landau}^{\mu\mu}=D_{0}k_{D}^{4} 2\pi \Lambda e^{-\frac{x^{2}}{2}}I_{0}\left (\frac{x^{2}}{2}\right ),
\label{reg11}
\end{equation} 
where $\Lambda=\int_{0}^{+\infty}{dk}/{k^{2}}$. We note that the trace
of the diffusion coefficient decays like $x^{-1}$ in each case. We can
write Eq. (\ref{reg10}) in the form of Eq. (\ref{reg11}):
\begin{equation}
D^{\mu\mu}=D_{0}k_{D}^{4} 2\pi \Lambda(x) e^{-\frac{x^{2}}{2}}I_{0}\left (\frac{x^{2}}{2}\right ),
\label{reg12}
\end{equation} 
where $\Lambda(x)$ is now a function of $x$ which is perfectly well defined (without divergence). Using Eq. (\ref{reg10}) and comparing with Eq. (\ref{reg12}) we find for $x\rightarrow 0$ that
\begin{equation}
\Lambda(0)=\frac{\pi}{4k_{D}}\simeq \frac{0.785...}{k_{D}},
\label{reg13}
\end{equation} 
and for $x\rightarrow +\infty$ that
\begin{equation}
\Lambda(+\infty)=\frac{\pi+\sqrt{2\pi}}{4k_{D}}\simeq \frac{1.412...}{k_{D}}.
\label{reg14}
\end{equation} 
This justifies, without introducing an {\it ad hoc} large-scale cut-off,  that $\Lambda$ is of order the Debye length $k_{D}^{-1}$ as expected. We also note that the function
\begin{equation}
k_{D}\Lambda(x)=\frac{1}{4}\left (\pi+\sqrt{2\pi}\frac{I_{1}(x^{2}/2)}{I_{0}(x^{2}/2)}\right ),
\label{reg15}
\end{equation} 
does not vary crucially with $x$ and remains of order unity (see
Fig. \ref{interpol}). Therefore, using the Landau approximation and
introducing a large scale cut-off at the Debye length $k_{D}^{-1}$
seems to be a reasonably good approximation. However, when one
evaluates the component $D_{\|}(x)$ of the diffusion coefficient, one
finds that the treatment using Eq. (\ref{reg6}) leads to a decay like
$x^{-2}$ for $x\rightarrow +\infty$ while the Landau approximation
(\ref{reg7}) leads to a decay like $x^{-3}$ (see
Sec. \ref{sec_wb}). Therefore, in that case, there are qualitative
discrepencies between the two approaches.

\end{document}